\pdfoutput=1

\documentclass[11pt,prd,aps,amssymb,amsmath,tightenlines,showpacs]{revtex4}
\usepackage{graphicx}

\newcommand{\cPT}{\ensuremath{\mathcal{PT}}}
\newcommand{\half}{\mbox{$\textstyle{\frac{1}{2}}$}}

\begin{document}
\title{Painlev\'e Transcendents and $\cPT$-Symmetric Hamiltonians} 

\author{Carl M. Bender$^{a,b}$}\email{cmb@wustl.edu}
\author{Javad Komijani$^a$}\email{jkomijani@physics.wustl.edu}
\affiliation{$^a$Department of Physics, Washington University, St. Louis, MO
63130, USA\\
$^b$Department of Mathematical Science, City University London,\\
$\,\,$ Northampton Square, London EC1V 0HB, UK}

\date{\today}

\begin{abstract}
Unstable separatrix solutions for the first and second Painlev\'e transcendents
are studied both numerically and analytically. For a fixed initial condition,
say $y(0)=0$, there is a discrete set of initial slopes $y'(0)=b_n$ that give
rise to separatrix solutions. Similarly, for a fixed initial slope, say $y'(0)=
0$, there is a discrete set of initial values $y(0)=c_n$ that give rise to
separatrix solutions. For Painlev\'e I the large-$n$ asymptotic behavior of
$b_n$ is $b_n\sim B_{\rm I}n^{3/5}$ and that of $c_n$ is $c_n\sim C_{\rm I}n^{2/
5}$, and for Painlev\'e II the large-$n$ asymptotic behavior of $b_n$ is $b_n
\sim B_{\rm II}n^{2/3}$ and that of $c_n$ is $c_n\sim C_{\rm II}n^{1/3}$. The
constants $B_{\rm I}$, $C_{\rm I}$, $B_{\rm II}$, and $C_{\rm II}$ are first
determined numerically. Then, they are found analytically and in closed form
by reducing the nonlinear equations to the linear eigenvalue problems associated
with the cubic and quartic $\cPT$-symmetric Hamiltonians $H=\half p^2+2ix^3$ and
$H=\half p^2-\half x^4$.
\end{abstract}

\maketitle

\section{Introduction}
\label{s1}
The famous Painlev\'e transcendents are six nonlinear second-order differential
equations whose key features are that their movable (spontaneous) singularities
are poles (and not, for example, branch points or essential singularities).
There is a vast literature on these remarkable differential equations
\cite{r1,r2,r3,r4,r5,r6,r7,r8}. These equations have arisen many times in
mathematical physics; for a small sample, see
Refs.~\cite{r9,r10,r11,r12,r13,r14,r15,r16}. This paper considers the first
and second Painlev\'e transcendents, referred to here as P-I and P-II. The
initial-value problem (IVP) for the P-I differential equation is
\begin{equation}
y''(t)=6[y(t)]^2+t,\qquad y(0)=c,~y'(0)=b
\label{e1}
\end{equation}
and the IVP for P-II (we have set an arbitrary additive constant to 0) is
\begin{equation}
y''(t)=2[y(t)]^3+ty(t),\qquad y(0)=c,~y'(0)=b.
\label{e2}
\end{equation}

Many asymptotic studies of the Painlev\'e transcendents have been published, but
in this paper we present a simple numerical and asymptotic analysis that to our
knowledge has not appeared in the literature. This analysis concerns the initial
conditions that give rise to special unstable separatrix solutions of P-I and
P-II. Our asymptotic analysis verifies the numerical results given in this paper
for P-I and P-II as well as some preliminary numerical calculations that were
presented in an earlier paper on nonlinear differential-equation eigenvalue
problems \cite{r17}.

The main idea, originally introduced in Ref.~\cite{r17}, is that a nonlinear
differential equation may have a discrete set of {\it critical} initial
conditions that give rise to unstable separatrix solutions. These discrete
initial conditions can be thought of as eigenvalues and the separatrices that
stem from these initial conditions can be viewed as the corresponding
eigenfunctions. The objective in Ref.~\cite{r17} was to find the large-$n$
(semiclassical) asymptotic behavior of the $n$th eigenvalue. The general
analytical approach that was proposed was to simplify the nonlinear differential
problem to a linear problem that could be used to determine the leading
asymptotic behavior of the eigenvalues as $n\to\infty$.

A toy model was used in Ref.~\cite{r17} to explain the concept of a nonlinear
eigenvalue problem. This model makes use of the elementary first-order
differential equation problem
\begin{equation}
y'(t)=\cos[\pi t\,y(t)],\quad y(0)=a.
\label{e3}
\end{equation}
It was shown that the solutions to this initial-value problem pass through $n$
maxima before vanishing like $1/t$ as $t\to\infty$. As the initial condition $a=
y(0)$ increases past special critical values $a_n$, the number of maxima jumps
from $n$ to $n+1$. At these critical values the solution $y(t)$ to (\ref{e3}) is
an unstable separatrix curve in the following sense: At values of $y(0)$
infinitesimally below $a_n$ the solution merges with a bundle of stable
solutions all having $n$ maxima and when $y(0)$ is infinitesimally above $a_n$
the solution merges with a bundle of stable solutions all having $n+1$ maxima.
The challenge is to determine the asymptotic behavior of the critical values
$a_n$ for large $n$. (This generic problem is the analog of a semiclassical
high-energy approximation in quantum mechanics.) To solve this problem it was
shown that for large $n$, the nonlinear differential equation problem (\ref{e3})
reduces to a {\it linear} one-dimensional random-walk problem. The random-walk
problem was solved exactly, and it was shown analytically that
\begin{equation}
a_n\sim 2^{5/6}\sqrt{n}\quad(n\to\infty).
\label{e4}
\end{equation}
Kerr subsequently found an alternative solution to this asymptotics problem and
verified (\ref{e4}) \cite{r18}.

The nonlinear eigenvalue problem described above is similar in many respects to
the linear eigenvalue problem for the time-independent Schr\"odinger equation.
For a potential $V(x)$ that rises as $x\to\pm\infty$, the eigenfunctions $\psi(
x)$ of the Schr\"odinger eigenvalue problem
\begin{equation}
-\psi''(x)+V(x)\psi(x)=E\psi(x),\quad\psi(\pm\infty)=0,
\label{e5}
\end{equation}
are unstable with respect to small changes in the eigenvalue $E$; that is, if
$E$ is increased or decreased slightly, $\psi(x)$ abruptly ceases to obey the
boundary conditions [and thus is not normalizable (square integrable)].
Furthermore, like the eigenfunctions (separatrix curves) of (\ref{e3}), the
eigenfunction $\psi_n(x)$ corresponding to the $n$th eigenvalue has $n$
oscillations in the classically allowed region before decreasing monotonically
to $0$ in the classically forbidden region.

This paper considers four eigenvalue problems. First, for P-I we find the
large-$n$ behavior of the positive eigenvalues $b_n$ for the initial condition
$y(0)=0,\,y'(0)=b_n$ and also the large-$n$ behavior of the negative eigenvalues
$c_n$ for the initial condition $y(0)=c_n,\,y'(0)=0$. We show that
$$b_n\sim B_{\rm I}n^{3/5}\quad{\rm and}\quad c_n\sim C_{\rm I}n^{2/5}.$$
Second, for P-II we show that for large $n$ the asymptotic behaviors of $b_n$
and $c_n$ are given by
$$b_n\sim B_{\rm II}n^{2/3}\quad{\rm and}\quad c_n\sim C_{\rm II}n^{1/3}.$$
We determine the constants $B_{\rm I}$, $C_{\rm I}$, $B_{\rm II}$, and $C_{\rm
II}$ both numerically and analytically.

This paper is organized as follows. In Sec.~\ref{s2} we obtain the constants
$B_{\rm I}$ and $C_{\rm I}$ by using numerical techniques and in Sec.~\ref{s3}
we do so analytically by reducing the large-eigenvalue problem to the {\it
linear} time-independent Schr\"odinger equation for the cubic $\cPT$-symmetric
Hamiltonian $H=\half p^2+ix^3$. Next, we study the eigenvalue problem for the
second Painlev\'e transcendent. In Sec.~\ref{s4} we present a numerical
determination of the large-$n$ behavior of the eigenvalues and in Sec.~\ref{s5}
we verify the numerical results in Sec.~\ref{s4} by using asymptotic analysis to
reduce the nonlinear large-eigenvalue problem for P-II to the linear
Schr\"odinger equation for the quartic $\cPT$-symmetric Hamiltonian $H=\half p^2
-\half x^4$. In Sec.~\ref{s6} we make some brief concluding remarks.

\section{Numerical analysis of the first Painlev\'e transcendent}
\label{s2}
In Ref.~\cite{r17} there is a brief numerical study of the initial-value problem
for the first Painlev\'e transcendent (\ref{e1}). It is easy to see that there
are two possible asymptotic behaviors as $t\to-\infty$; the solutions to the P-I
equation can approach either $+\sqrt{-t/6}$ or $-\sqrt{-t/6}$. An elementary
asymptotic analysis shows that if the solution $y(t)$ approaches $-\sqrt{-t/6}$,
the solution oscillates stably about this curve with slowly decreasing amplitude
\cite{r19}. However, while the curve $+\sqrt{-t/6}$ is a possible asymptotic
behavior, this behavior is {\it unstable} and nearby solutions tend to veer away
from it. We define the eigenfunction solutions to the first Painlev\'e
transcendent as those solutions that {\it do} approach $+\sqrt{-t/6}$ as $t\to-
\infty$. These separatrix solutions resemble the eigenfunctions of conventional
quantum mechanics in that they exhibit $n$ oscillations before settling down to
this asymptotic behavior. However, because the P-I equation is nonlinear, these
oscillations are violent; the $n$th eigenfunction passes through $[n/2]$ double
poles where it blows up before it smoothly approaches the curve $+\sqrt{-t/6}$.
(The symbol $[n/2]$ means greatest integer in $n/2$.)

One can specify two different kinds of eigenvalue problems for P-I, each of
which is fundamentally related to the instability of the asymptotic behavior
$+\sqrt{-t/6}$. One can (i) fix the initial value $y(0)$ and look for (discrete)
values of the initial slopes $y'(0)=b$ that give rise to solutions approaching
$+\sqrt{-t/6}$, or else (ii) one can fix the initial slope $y'(0)$ and look for
the (discrete) initial values of $y(0)=c$ that give rise to solutions
approaching $+\sqrt{-t/6}$.

\subsection{Initial-slope eigenvalues for Painlev\'e I}
\label{ss2a}
Let us examine the numerical solutions to the initial-value problem for the P-I
equation (\ref{e1}) for $t<0$. To find these solutions we use Runge-Kutta to
integrate down the negative-real axis. When we approach a double pole and the
solution becomes large and positive, we estimate the location of the pole and
integrate along a semicircle in the complex-$t$ plane around the pole. We then
continue integrating down the negative-real axis. We choose the fixed initial
value $y(0)=0$ and allow the initial slope $y'(0)=b$ to have increasingly
positive values. (We only present results for positive initial slope; the
behavior for negative initial slope is analogous and describing it would be
repetitive.) Our numerical analysis shows that the particular choice of $y(0)$
is not crucial; for {\it any} fixed $y(0)$ the large-$n$ asymptotic behavior of
the initial-slope eigenvalues $b_n$ is the same.

\begin{figure}[h!]
\null\vspace{-9mm}
\begin{center}
\includegraphics[trim=21mm 50mm 18mm 50mm,clip=true,scale=0.45]{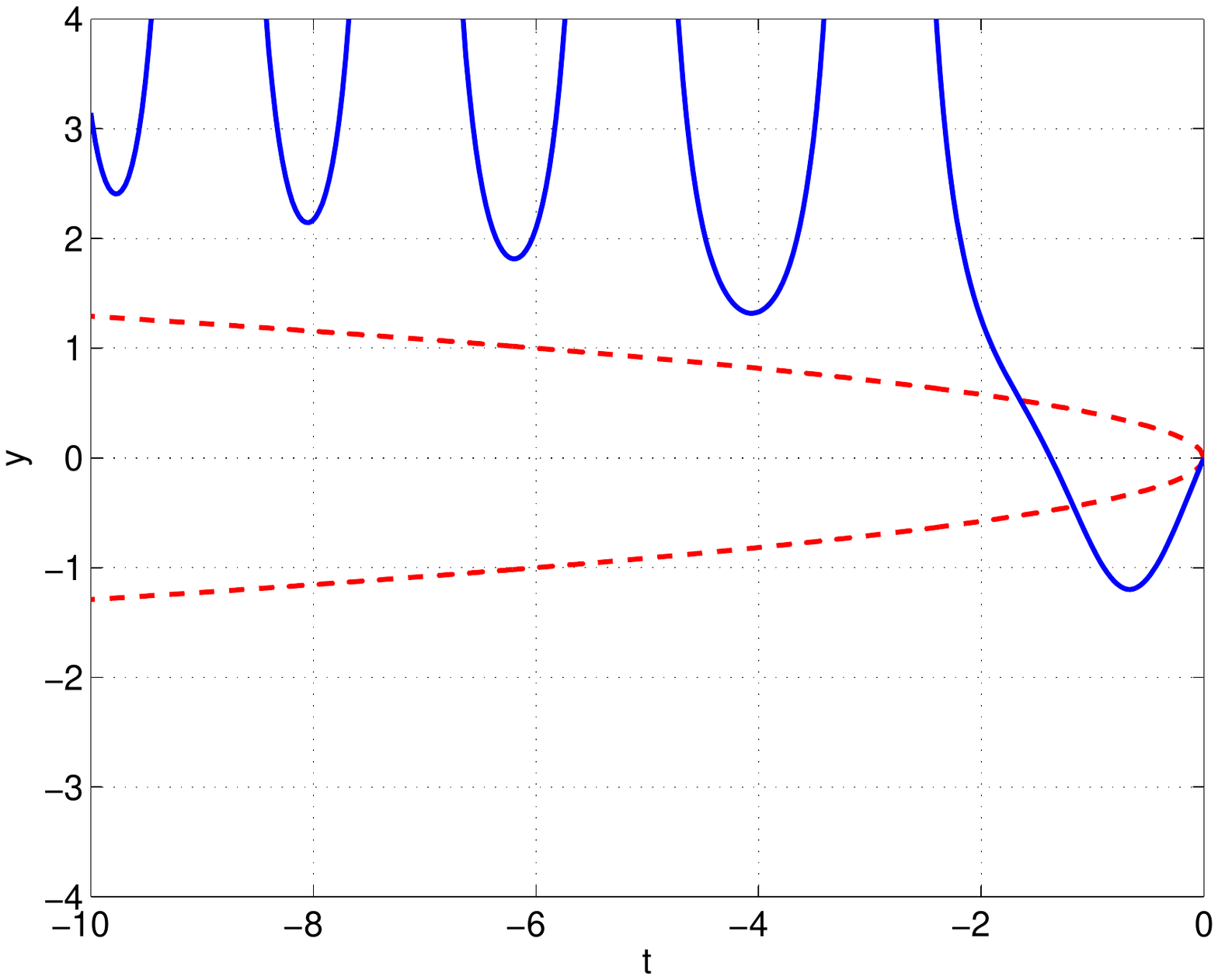}
% trim=left bottom right top
\includegraphics[trim=18mm 50mm 21mm 50mm,clip=true,scale=0.45]{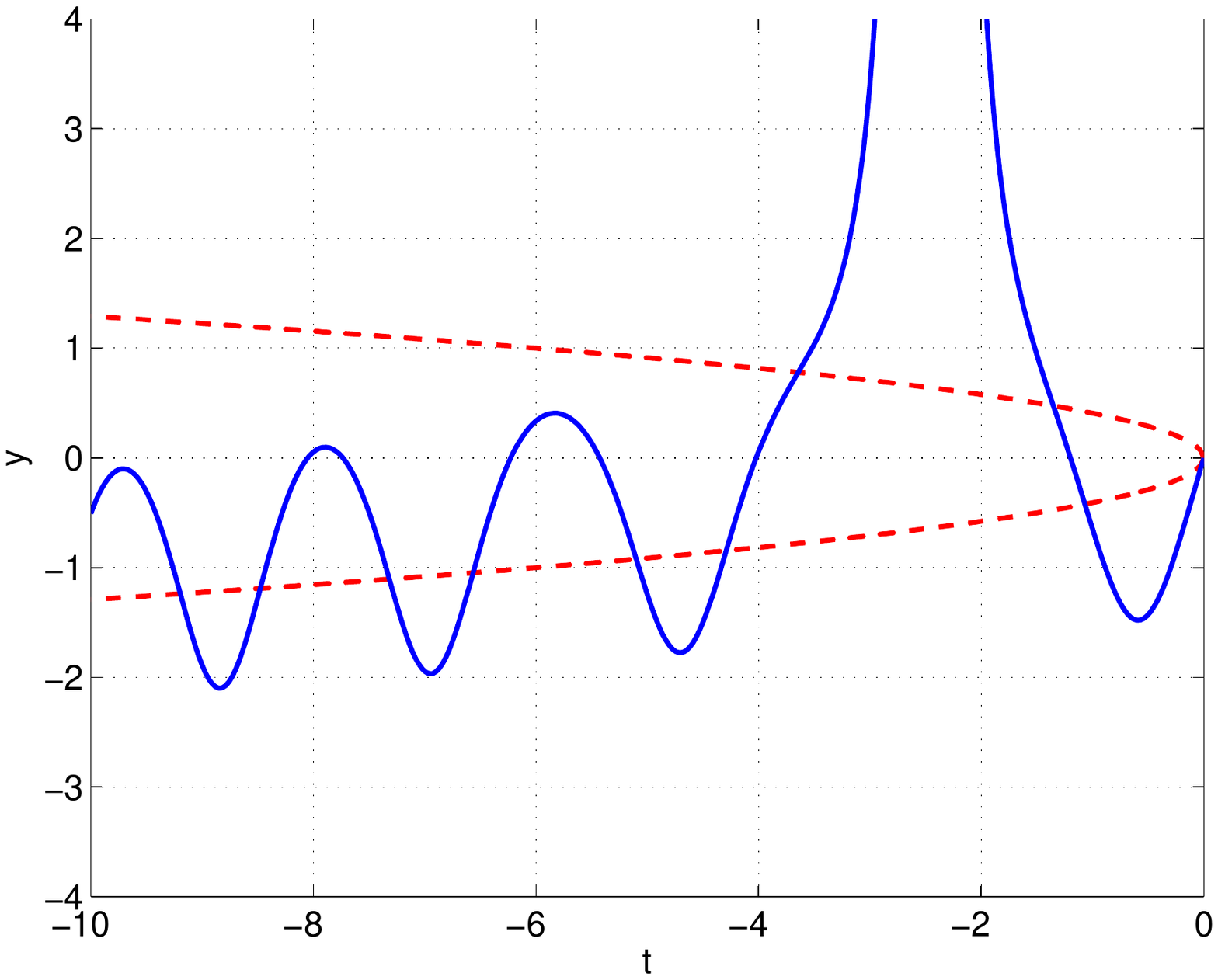}
\hspace{1.5cm}
\end{center}
\null\vspace{-21mm}
\caption{Typical behavior of solutions to the first Painlev\'e transcendent $y(t
)$ for the initial conditions $y(0)=0$ and $b=y'(0)$. In the left panel $b=
2.504031103$, which lies between the eigenvalues $b_1=1.851854034$ and $b_2=
3.004031103$. In the right panel $b=3.504031103$, which lies between the
eigenvalues $b_2=3.004031103$ and $b_3=3.905175320$. The dashed curves are $y=
\pm\sqrt{-t/6}$. In the left panel the solution $y(t)$ has an infinite sequence
of double poles and in the right panel the solution oscillates stably about
$-\sqrt{t/6}$.}
\label{F1}
\end{figure}

\begin{figure}[h!]
\null\vspace{-9mm}
\begin{center}
\includegraphics[trim=21mm 50mm 18mm 50mm,clip=true,scale=0.45]{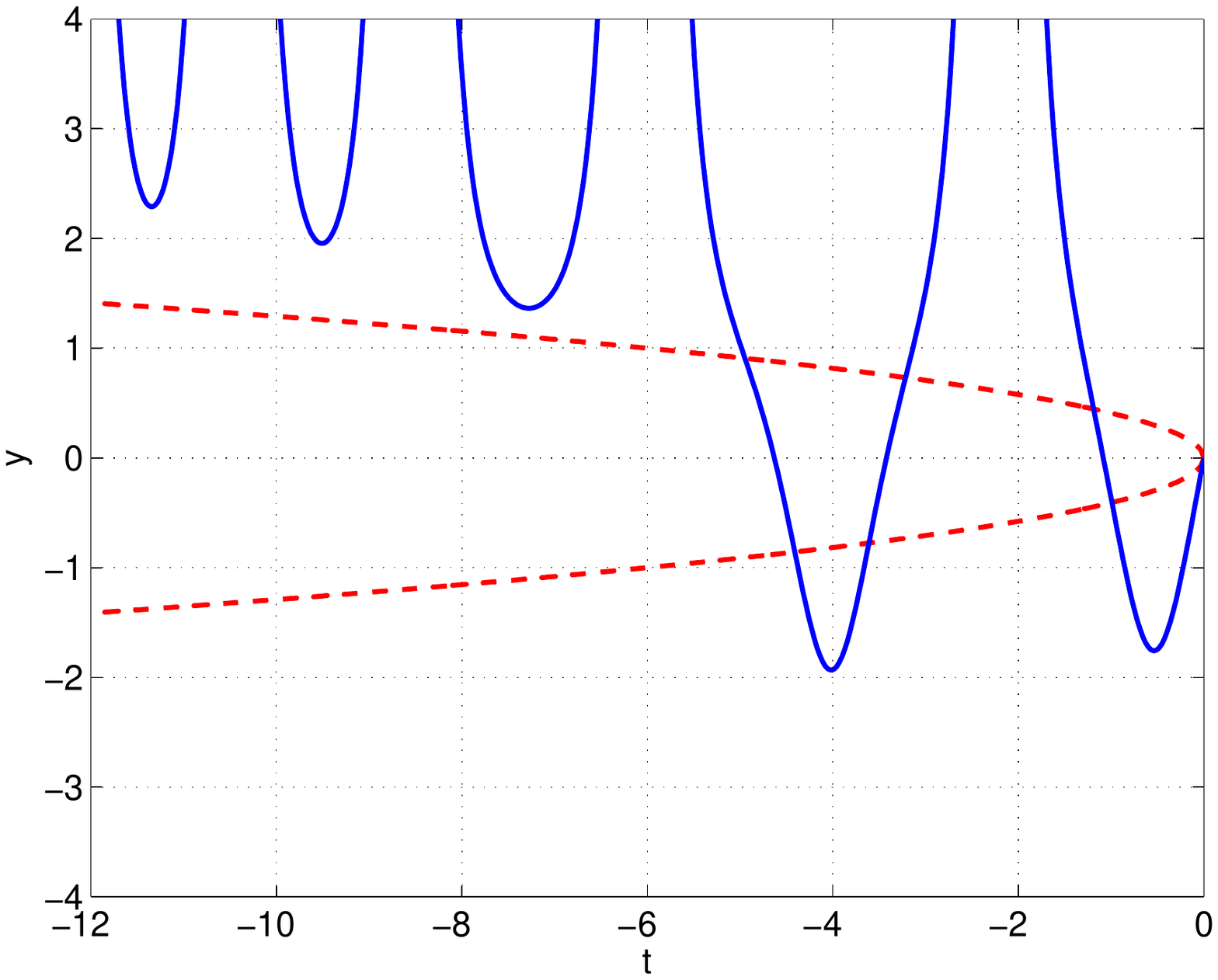}
% trim=left bottom right top
\includegraphics[trim=18mm 50mm 21mm 50mm,clip=true,scale=0.45]{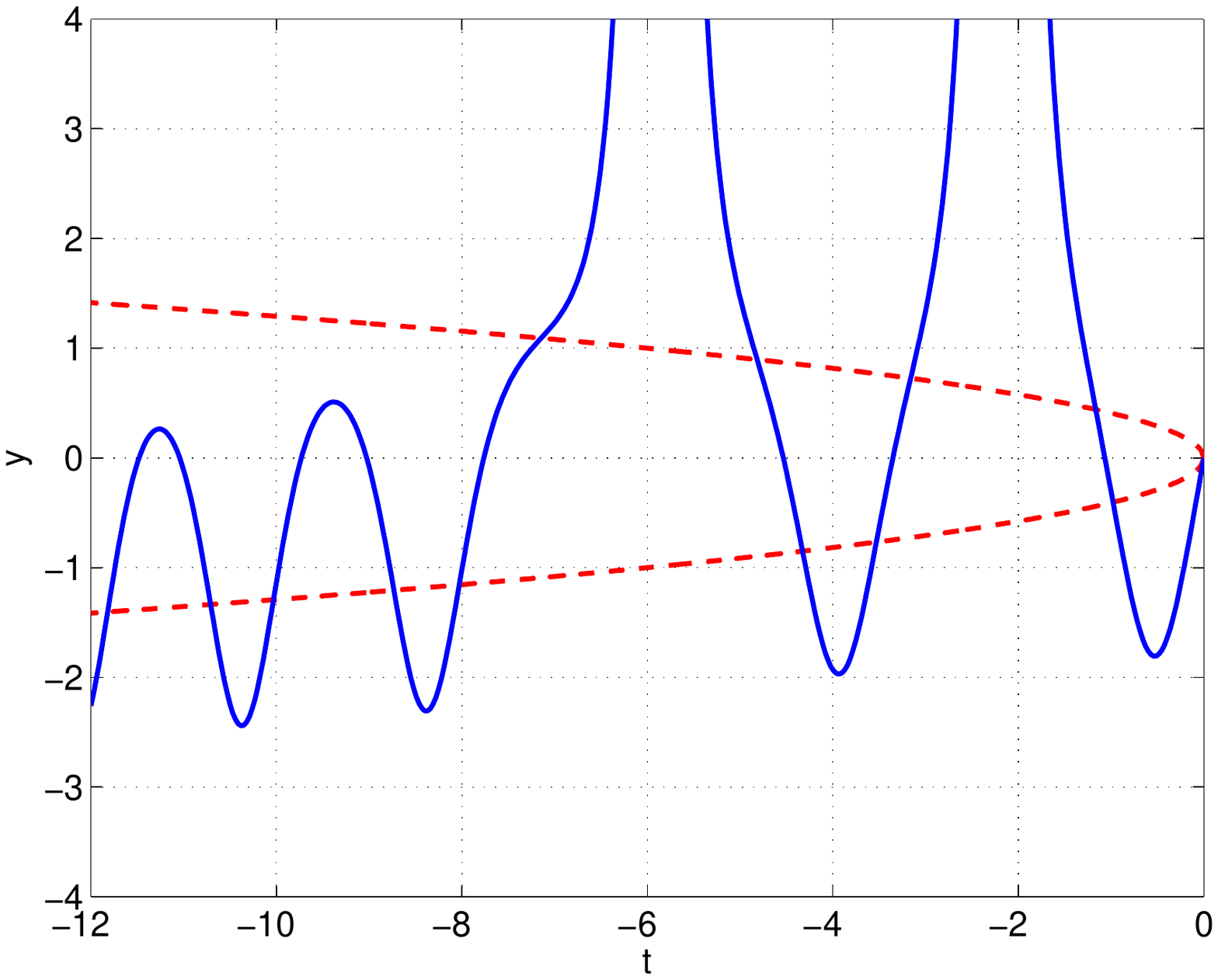}
\hspace{1.5cm}
\end{center}
\null\vspace{-21mm}
\caption{Solutions to the P-I equation (\ref{e1}) for $y(0)=0$ and $b=y'(0)$.
Left panel: $b=4.583412410$, which lies between the eigenvalues $b_3=
3.905175320$ and $b_4=4.6834124103$. Right panel: $b=4.783412410$, which
lies between the eigenvalues $b_4=4.683412410$ and $b_5=5.383086722$.}
\label{F2}
\end{figure}

We find that above the critical value $b_1=1.851854034$ (the first eigenvalue)
there is a continuous interval of $b$ for which $y(t)$ first has a minimum and
then has an infinite sequence of double poles (see Fig.~\ref{F1}, left panel).
However, if $b$ increases past the next critical value $b_2=3.004031103$ (the
second eigenvalue), the character of the solutions changes abruptly and $y(t)$
oscillates stably about $-\sqrt{-t/6}$ (Fig.~\ref{F1}, right panel). When $b$
exceeds the critical value $b_3=3.905175320$ (the third eigenvalue), the
solutions again exhibit an infinite sequence of poles (Fig.~\ref{F2}, left
panel). When $b$ increases past the fourth critical value $b_4=4.683412410$
(fourth eigenvalue), the solutions once again oscillate stably about $-\sqrt{-
t/6}$ (Fig.~\ref{F2}, right panel). Our numerical analysis indicates that there
is an infinite sequence of critical points (eigenvalues) at which the P-I
solutions alternate between infinite sequences of double poles and stable
oscillation about $-\sqrt{-t/6}$.

\begin{figure}[h!]
\null\vspace{-9mm}
\begin{center}
\includegraphics[trim=21mm 50mm 18mm 50mm,clip=true,scale=0.45]{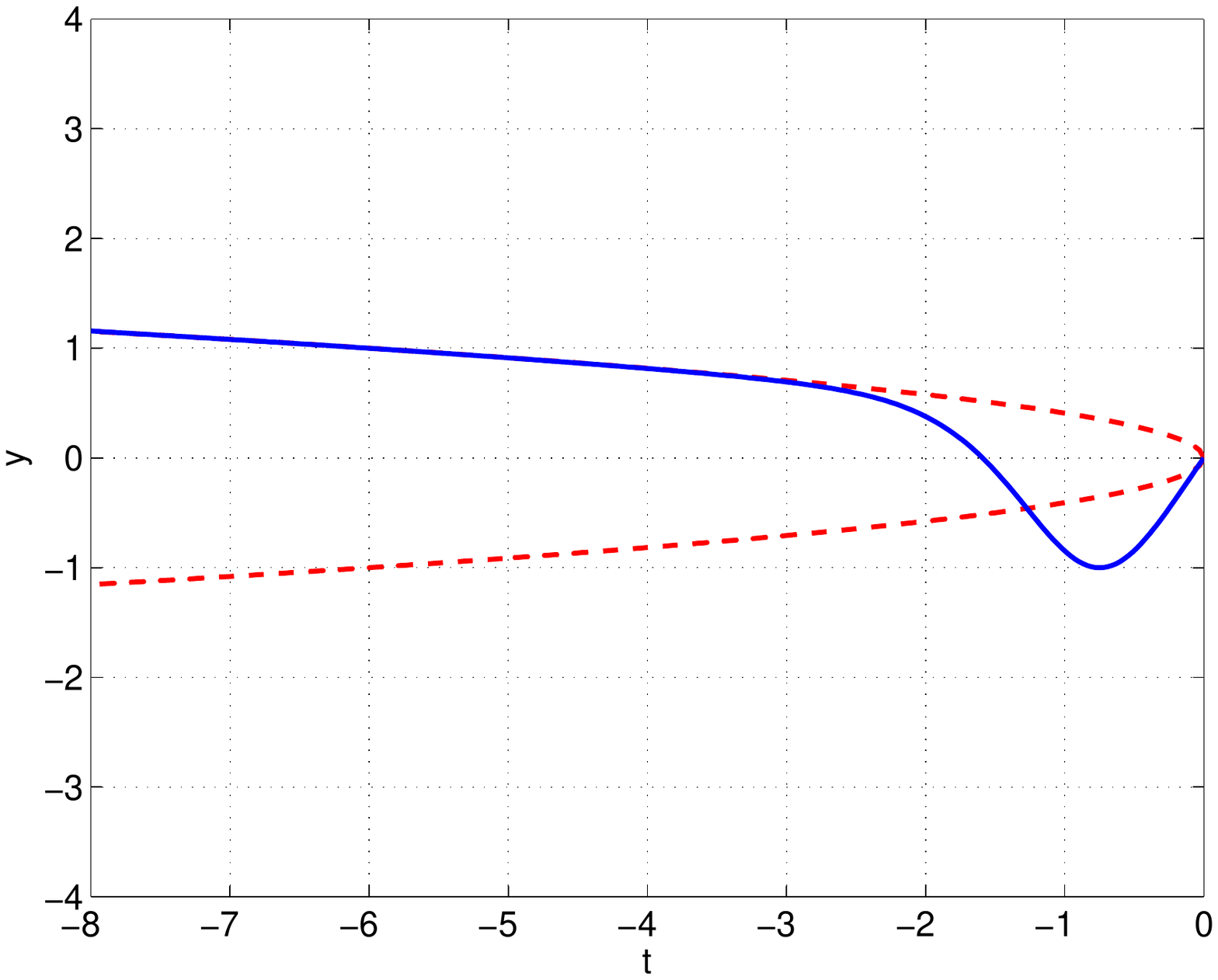}
% trim=left bottom right top
\includegraphics[trim=18mm 50mm 21mm 50mm,clip=true,scale=0.45]{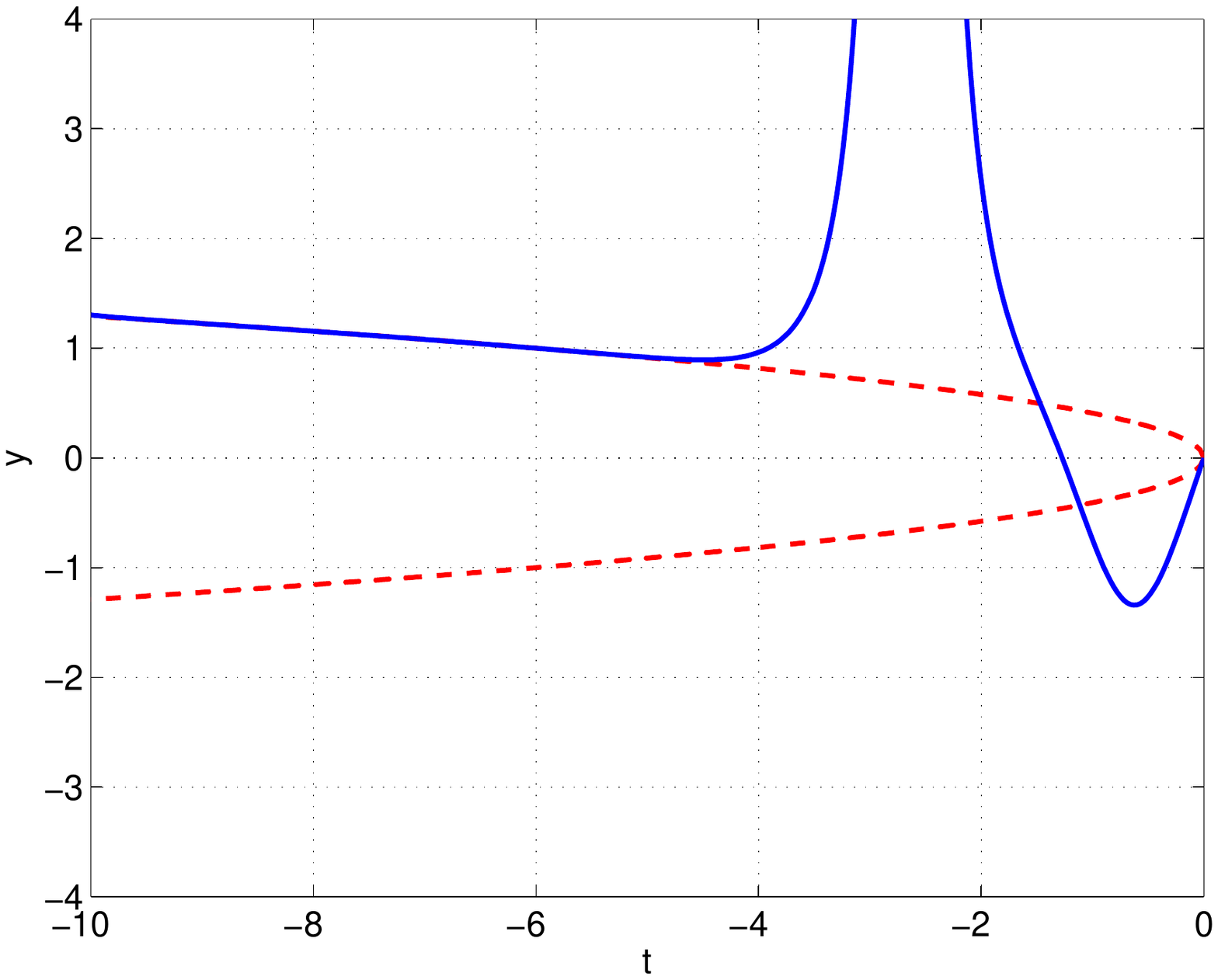}
\hspace{1.5cm}
\end{center}
\null\vspace{-21mm}
\caption{First two separatrix solutions (eigenfunctions) of Painlev\'e I with
initial condition $y(0)=0$. Left panel: $y'(0)=b_1=1.851854034$; right panel:
$y'(0)=b_2=3.004031103$. The dashed curves are $y=\pm\sqrt{-t/6}$.}
\label{F3}
\end{figure}

\begin{figure}[h!]
\null\vspace{-9mm}
\begin{center}
\includegraphics[trim=21mm 50mm 18mm 50mm,clip=true,scale=0.45]{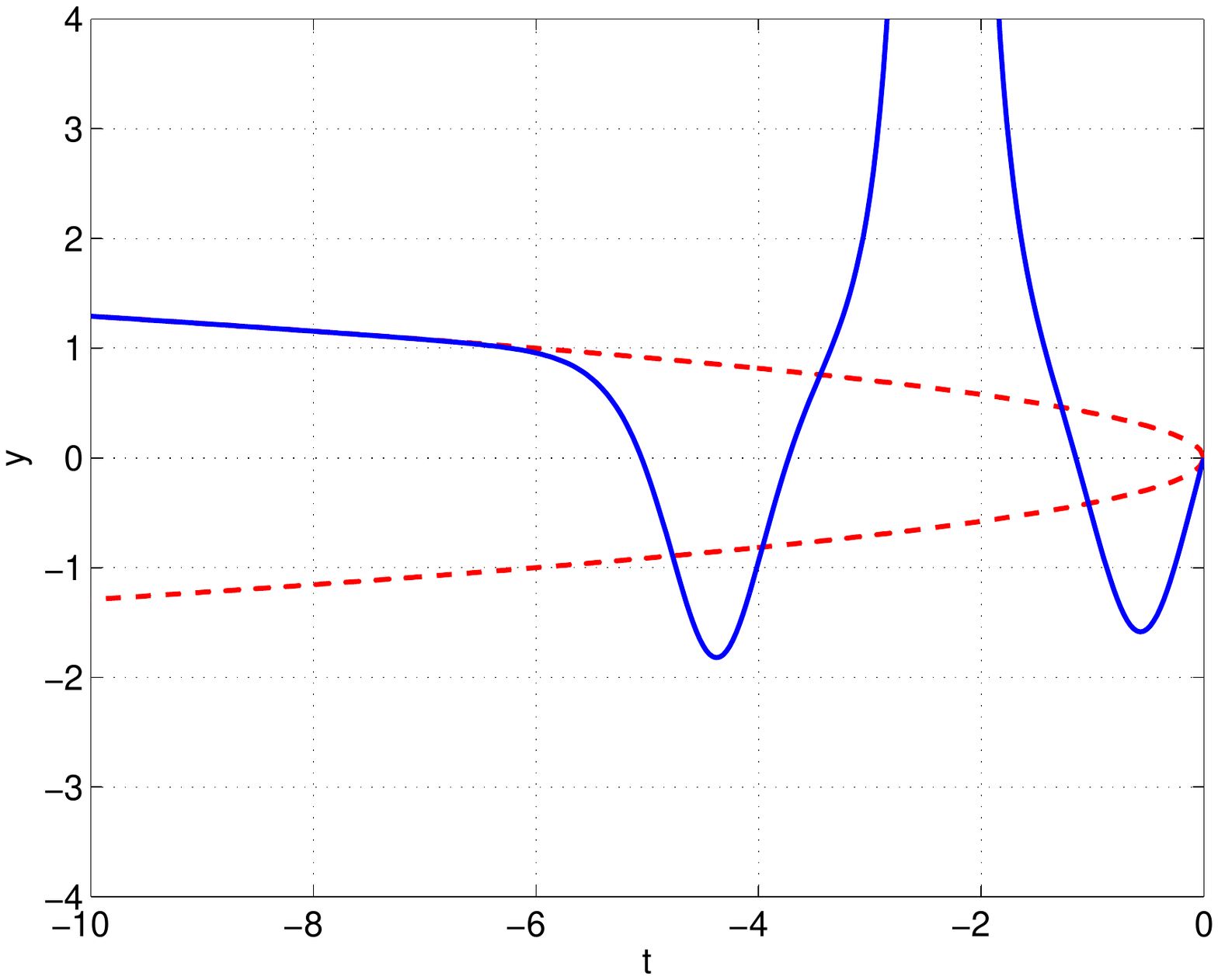}
% trim=left bottom right top
\includegraphics[trim=18mm 50mm 21mm 50mm,clip=true,scale=0.45]{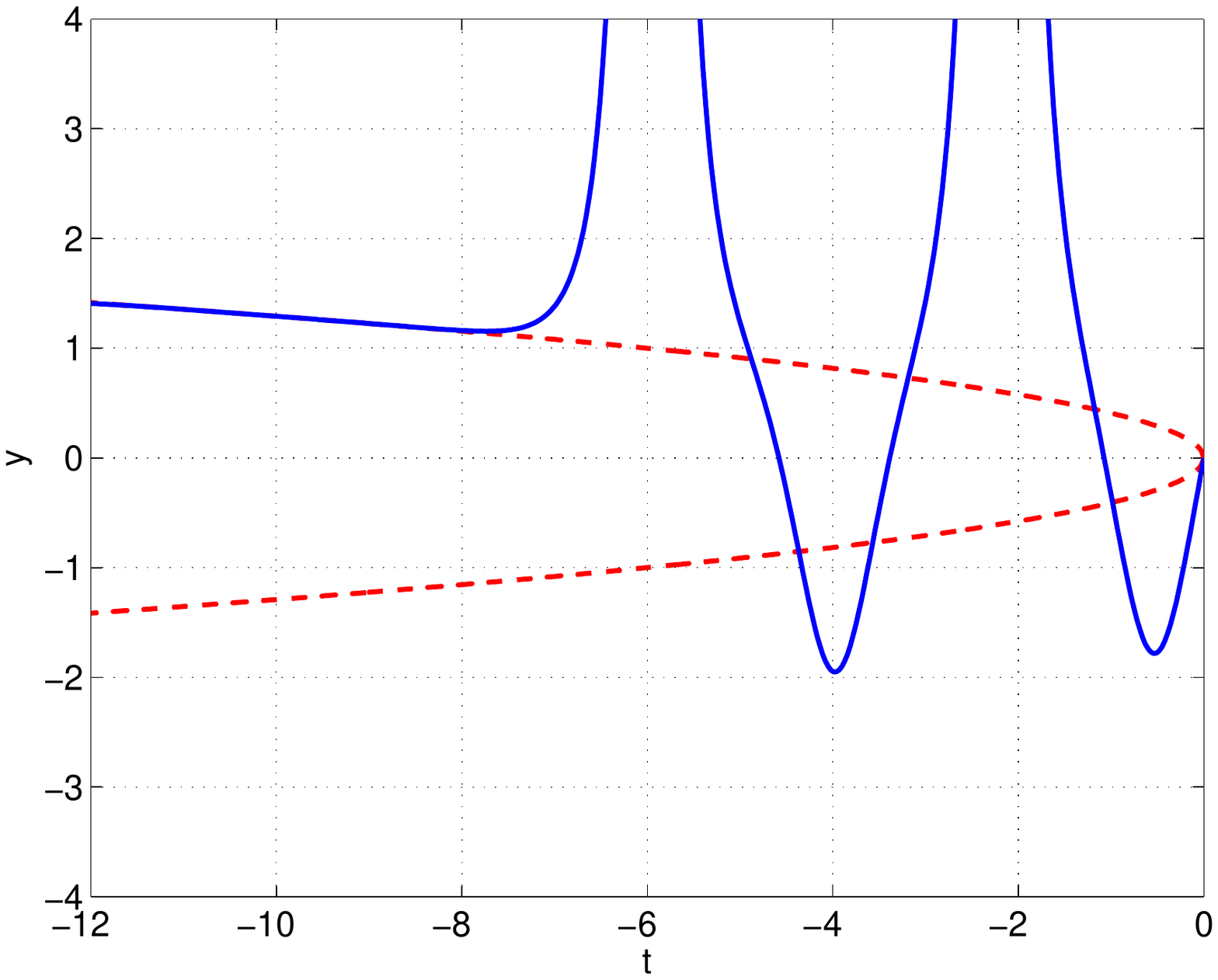}
\hspace{1.5cm}
\end{center}
\null\vspace{-21mm}
\caption{Third and fourth eigenfunctions of Painlev\'e I with initial condition
$y(0)=0$. Left panel: $y'(0)=b_3=3.905175320$; right panel: $y'(0)=b_4=
4.683412410$.}
\label{F4}
\end{figure}

\begin{figure}[h!]
\null\vspace{-9mm}
\begin{center}
\includegraphics[trim=21mm 50mm 18mm 50mm,clip=true,scale=0.45]{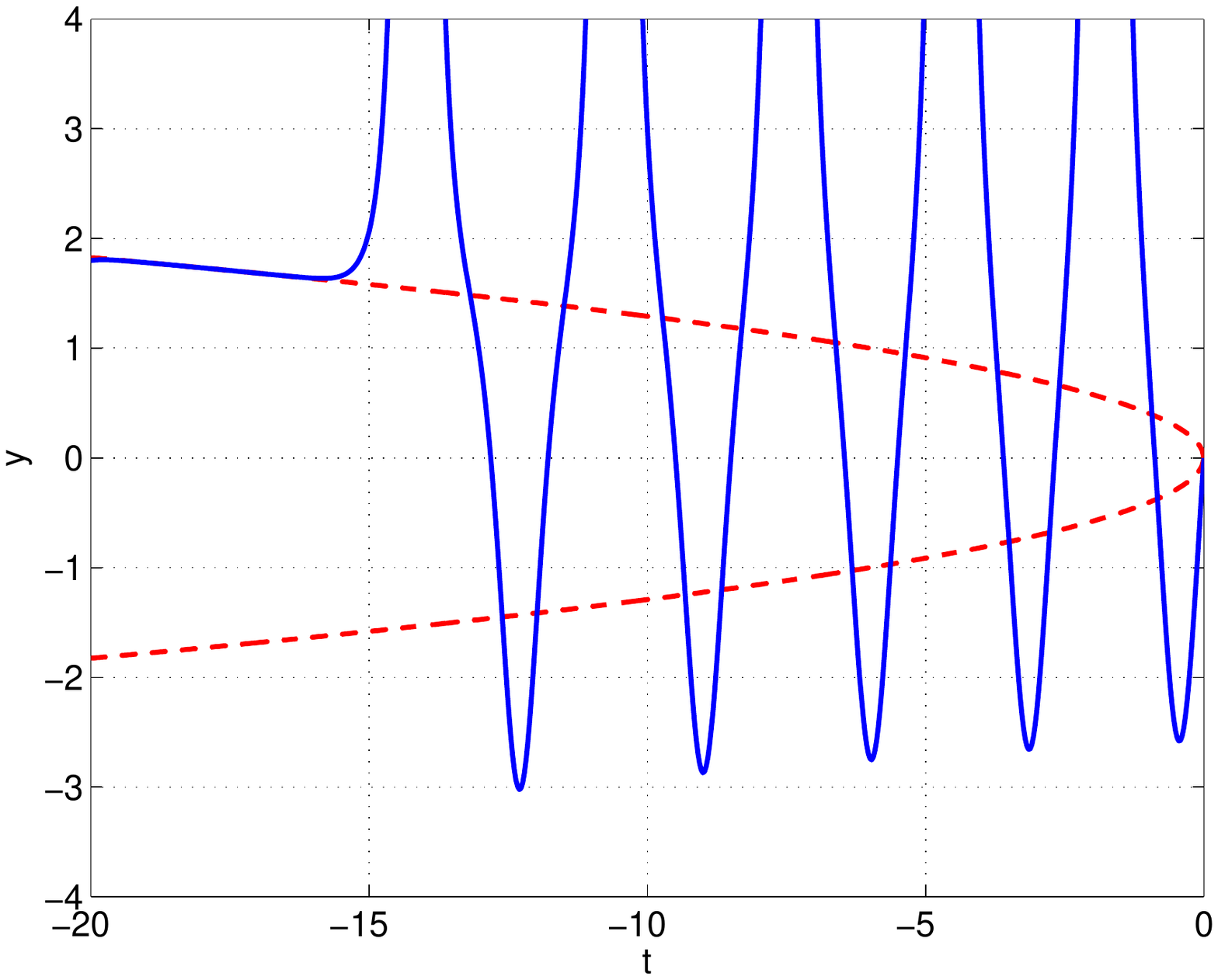}
% trim=left bottom right top
\includegraphics[trim=18mm 50mm 21mm 50mm,clip=true,scale=0.45]{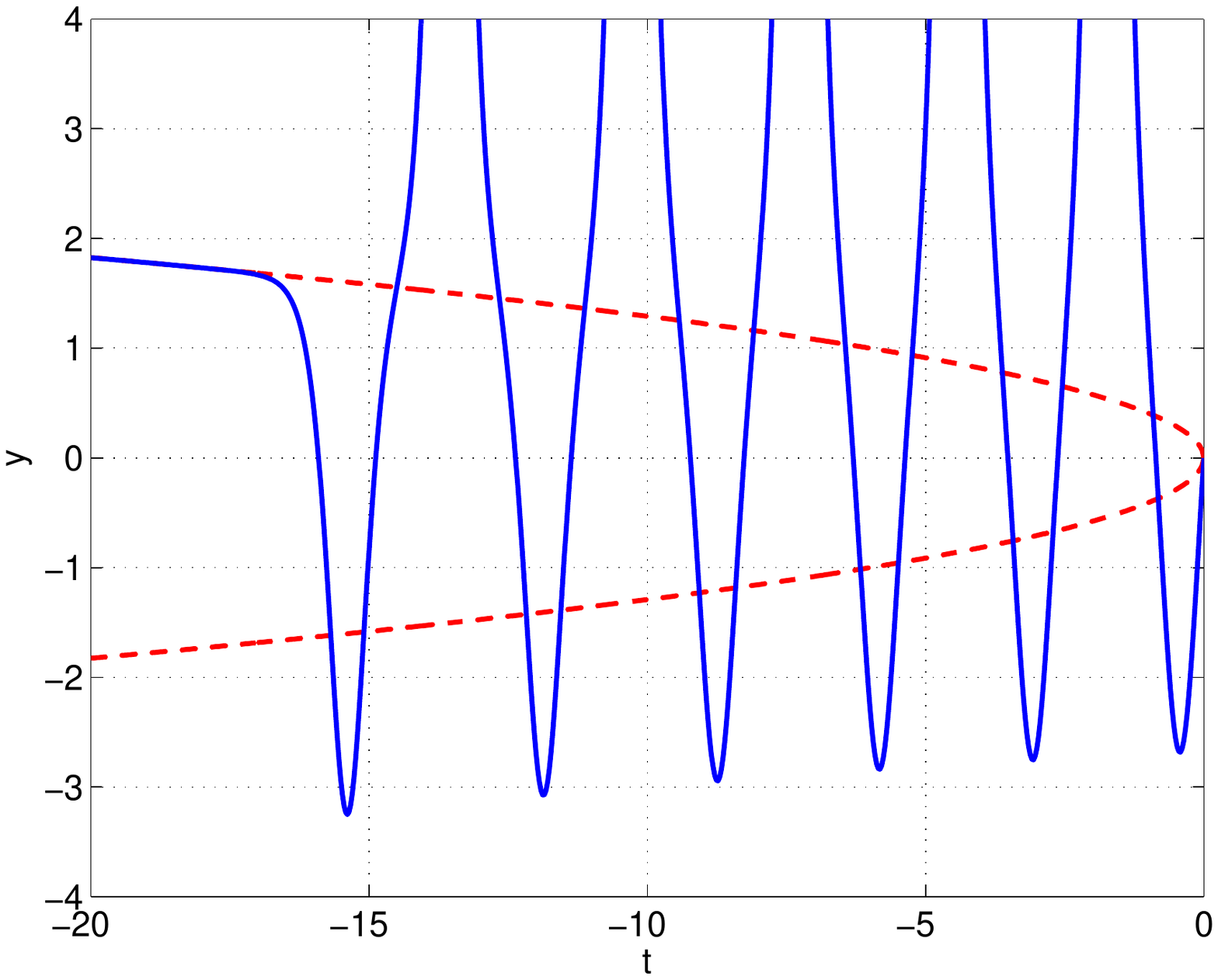}
\hspace{1.5cm}
\end{center}
\null\vspace{-21mm}
\caption{Tenth and eleventh eigenfunctions of Painlev\'e I with initial
condition $y(0)=0$. Left panel: $y'(0)=b_{10}=8.244932302$; right panel: $y'(0)=
b_{11}=8.738330156$. Note that as $n$ increases, the eigenfunctions pass through
more and more double poles before exhibiting a turning-point-like transition
and approaching the limiting curve $+\sqrt{-t/6}$ exponentially rapidly. This
behavior is analogous to that of the eigenfunctions of a time-independent
Schr\"odinger equation for a particle in a potential well; the higher-energy
eigenfunctions exhibit more and more oscillations in the classically allowed
region before entering the classically forbidden region, where they decay
exponentially.}
\label{F5}
\end{figure}

The solutions that arise when $y'(0)$ is at an eigenvalue have a completely
different (and unstable) character from those in Figs.~\ref{F1} and \ref{F2}.
These special solutions pass through a {\it finite} number of double poles
(analogous to the oscillatory behavior of quantum-mechanical bound-state
eigenfunctions in the classically allowed region of a potential well) and then
undergo a turning-point-like transition in which the poles cease and $y(t)$
exponentially approaches the limiting curve $+\sqrt{-t/6}$. The solutions
arising from the first and second critical points $b_1$ and $b_2$ are shown in
Fig.~\ref{F3}, those arising from the third and fourth critical points $b_3$ and
$b_4$ are shown in Fig.~\ref{F4}, and those arising from the tenth and eleventh
critical points $b_{10}$ and $b_{11}$ are shown in Fig.~\ref{F5}. The critical
points are analogous to eigenvalues because they give rise to {\it unstable}
separatrix solutions; if $y'(0)$ changes by an infinitesimal amount above or
below a critical value, the character of the solutions changes abruptly and the
solutions exhibit the two possible generic behaviors shown in Figs.~\ref{F1} and
\ref{F2}.

In Ref.~\cite{r17} a numerical asymptotic study of the critical values $b_n$ for
$n\gg1$ was performed by using Richardson extrapolation \cite{r20}. [In
Ref.~\cite{r17} the initial value was chosen to be $y(0)=1$ rather than $y(0)=0$
as in the current paper. However, as emphasized above, if $y(0)$ is held fixed,
we find that the large-$n$ asymptotic behavior of the initial slope $b_n$ is
insensitive to the choice of $y(0)$.] It was found in Ref.~\cite{r17} that for
large $n$, the $n$th critical value had the asymptotic behavior
\begin{equation}
y_n'(0)=b_n\sim B_{\rm I} n^{3/5}\quad(n\to\infty).
\label{e6}
\end{equation}
In Ref.~\cite{r17} the constant $B_{\rm I}$ was determined numerically to an
accuracy of about four or five decimal places. However, we have now performed a
more accurate numerical determination of the constant $B_{\rm I}$ by applying
fifth-order Richardson extrapolation to the first eleven eigenvalues, and we
have found the value of $B_{\rm I}$ accurate to one part in nine decimal places:
\begin{equation}
B_{\rm I}=2.0921467{\underbar 4}.
\label{e7}
\end{equation}
On the basis of our numerical analysis, we can say with confidence that the
underlined digit lies in the range from 3 to 5, so our determination of $B_{\rm
I}$ is accurate to one part in $2\times 10^8$.

\subsection{Initial-value eigenvalues for Painlev\'e I}
\label{ss2b}

\begin{figure}[h!]
\null\vspace{-9mm}
\begin{center}
\includegraphics[trim=21mm 50mm 18mm 50mm,clip=true,scale=0.45]{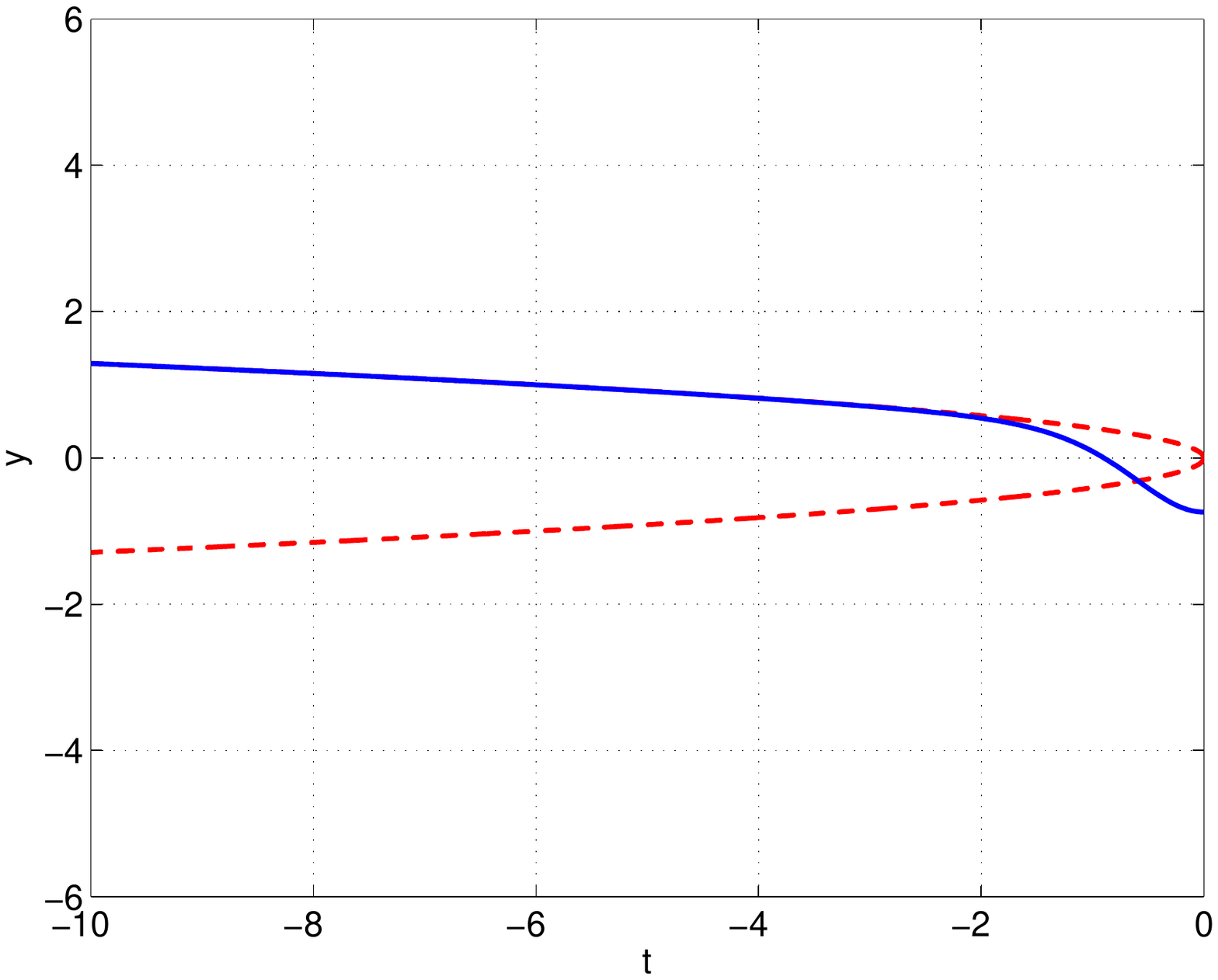}
% trim=left bottom right top
\includegraphics[trim=18mm 50mm 21mm 50mm,clip=true,scale=0.45]{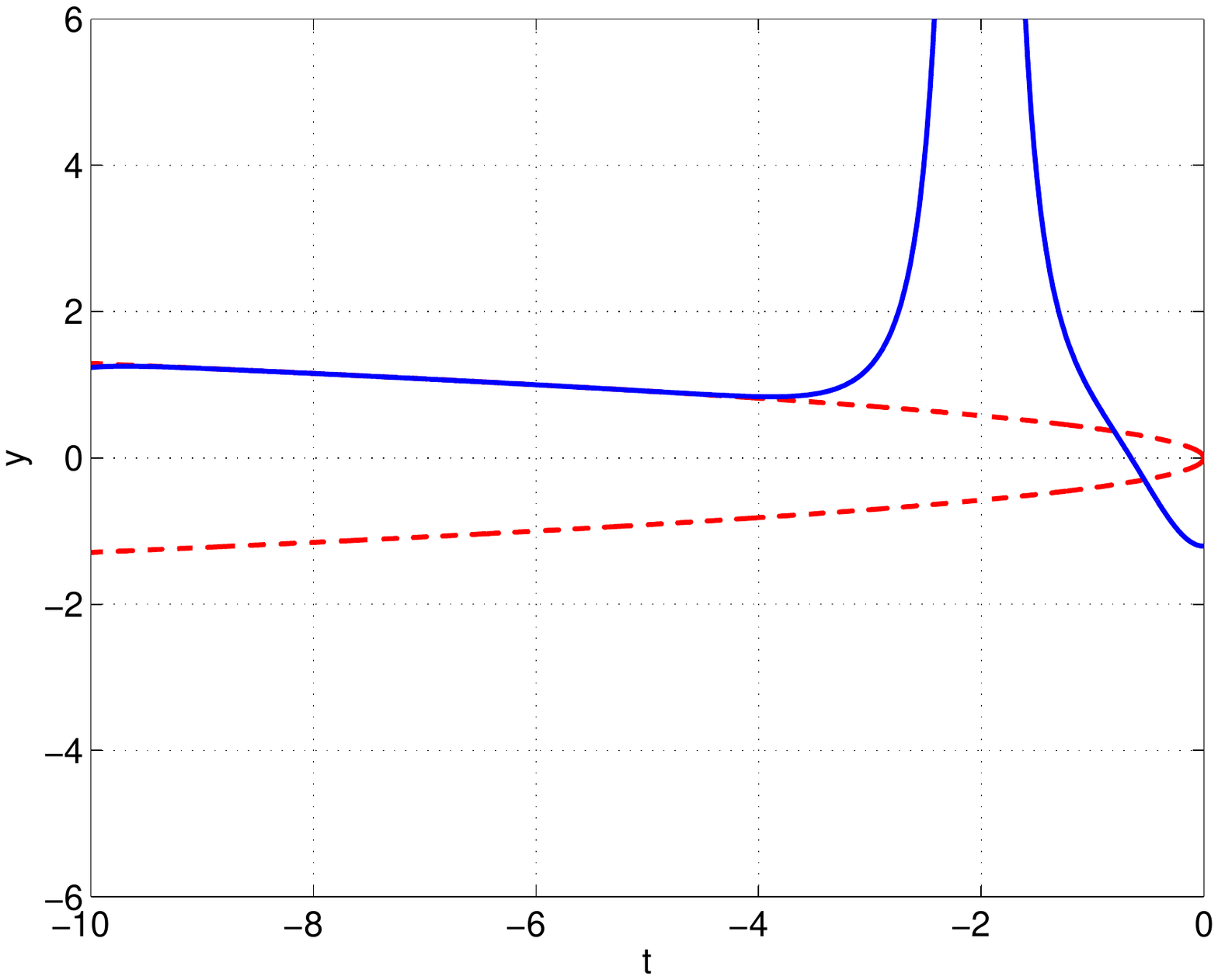}
\hspace{1.5cm}
\end{center}
\null\vspace{-21mm}
\caption{First two separatrix solutions (eigenfunctions) of Painlev\'e I with
fixed initial slope $y'(0)=0$. Left panel: $y(0)=c_1=-0.7401954236$; right
panel: $y(0)=c_2=-1.206703845$. The dashed curves are $y=\pm\sqrt{-t/6}$.}
\label{F6}
\end{figure}

\begin{figure}[h!]
\null\vspace{-9mm}
\begin{center}
\includegraphics[trim=21mm 50mm 18mm 50mm,clip=true,scale=0.45]{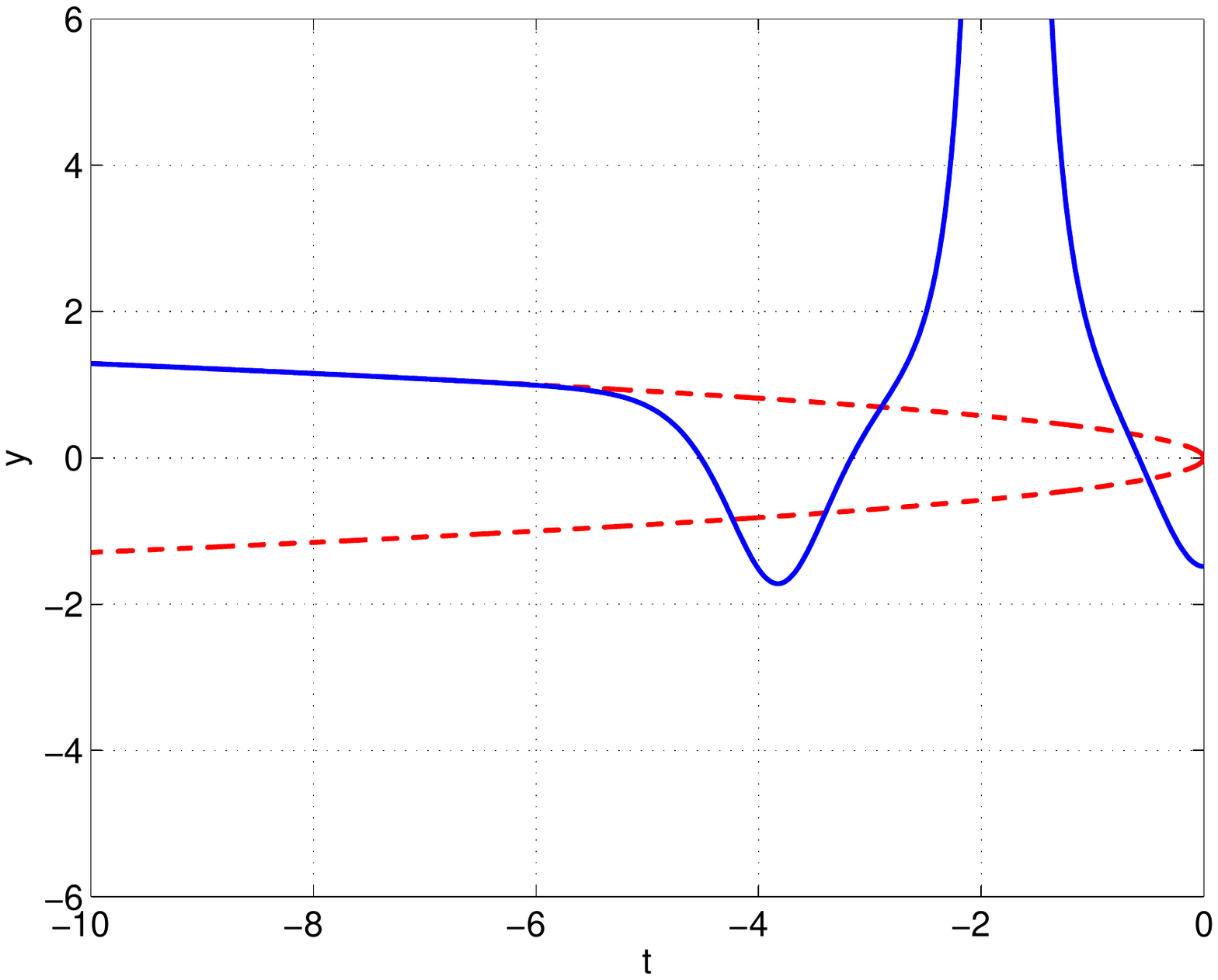}
% trim=left bottom right top
\includegraphics[trim=18mm 50mm 21mm 50mm,clip=true,scale=0.45]{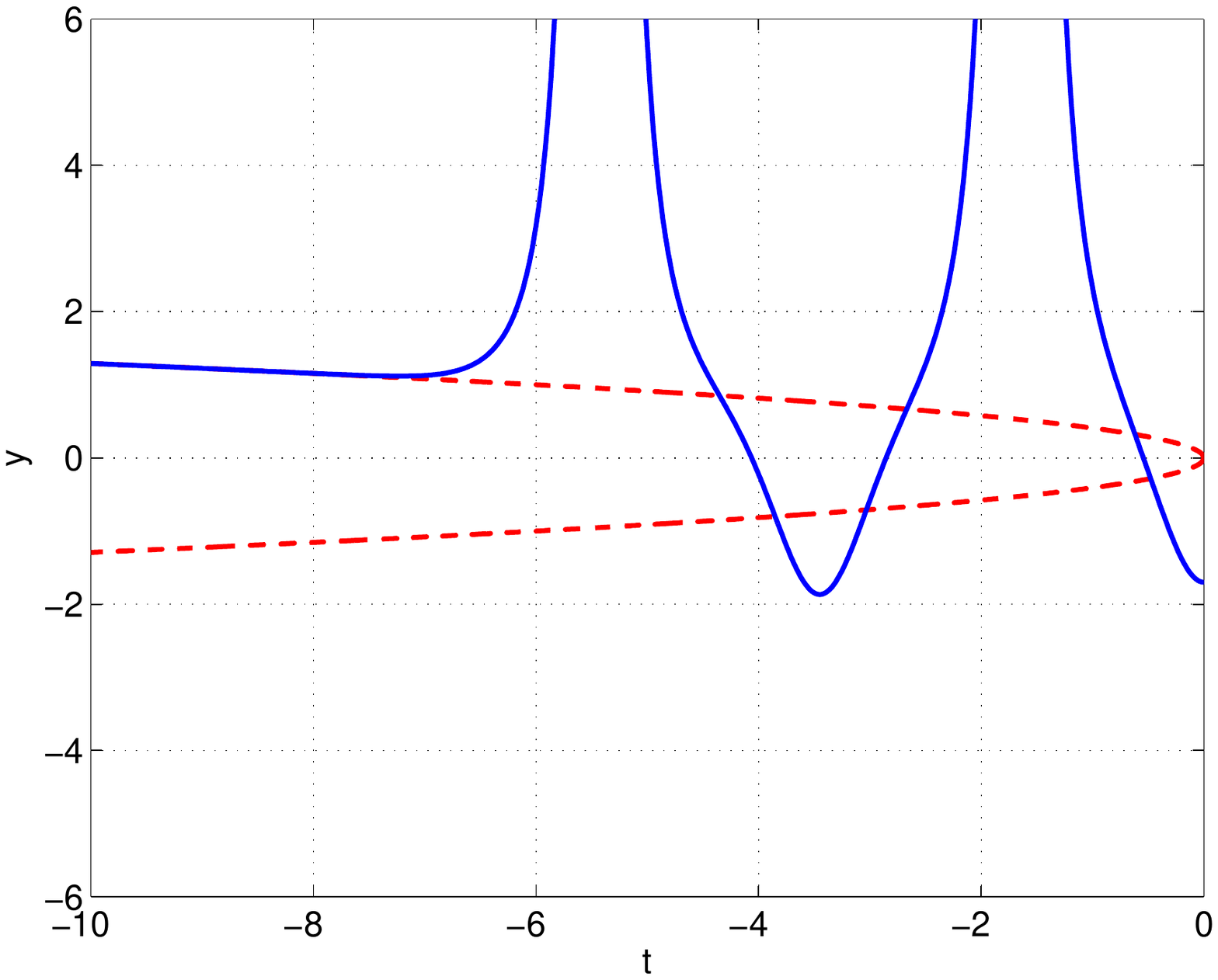}
\hspace{1.5cm}
\end{center}
\null\vspace{-21mm}
\caption{Third and fourth eigenfunctions of Painlev\'e I with initial slope $y'(
0)=0$. Left panel: $y(0)=c_3=-1.484375587$; right panel: $y(0)=c_4=
-1.69951765$.}
\label{F7}
\end{figure}

If we hold the initial slope fixed at $y'(0)=0$ and allow the initial value
$y(0)=c$ to become increasingly negative, we find that there is a sequence of
negative eigenvalues $c_n$ for which the solutions behave like the eigenfunction
separatrix solutions in Figs.~\ref{F3}--\ref{F5}. The first four eigenfunctions
are plotted in Figs.~\ref{F6} and \ref{F7}.

Applying fourth-order Richardson extrapolation to the first 15 eigenvalues, we
find that for large $n$ the sequence of initial-value eigenvalues $c_n$ is
asymptotic to $C_{\rm I}n^{2/5}$, where the numerical value of the constant
$C_{\rm I}$ is
\begin{equation}
C_{\rm I}=-1.030484{\underbar 4}.
\label{e8}
\end{equation}
We are confident that the final digit is accurate to an error of $\pm1$ and thus
$C_{\rm I}$ is determined to an accuracy of one part in $10^7$.

\section{Asymptotic calculation of $B_{\rm I}$ and $C_{\rm I}$}
\label{s3}
In this section we present an analytic calculation of $B_{\rm I}$ and $C_{\rm
I}$ in (\ref{e7}) and (\ref{e8}). To begin, we multiply the P-I differential
equation in (\ref{e1}) by $y'(t)$ and integrate from $t=0$ to $t=x$. We get
\begin{equation}
H\equiv\half[y'(x)]^2-2[y(x)]^3=\half[y'(0)]^2-2[y(0)]^3+I(x),
\label{e9}
\end{equation}
where $I(x)=\int_0^x dt\,ty'(t)$. Note that the path of integration is the same
as that used to calculate $y(t)$ numerically in Sec.~\ref{s2}; it follows the
negative-real axis until is gets near a pole, at which point it makes a
semicircular detour in the complex-$t$ plane to avoid the pole.

If we evaluate $I(x)$ for large $|x|$ in the classically allowed region (just
before the poles abruptly cease at the turning point), we find that as $n\to
\infty$, $I(x)$ fluctuates and becomes small compared with $H$. This is not
surprising because $I(x)$ receives many positive and negative contributions
from the poles. [In fact, by calculating $I(x)$ as $x\to-\infty$, we can see a
clear signal of an eigenvalue; as $y'(0)=b$ passes an eigenvalue, $I(x)$ goes
from having positive to negative (or negative to positive) fluctuations but at
an eigenvalue $I(x)$ is smooth and not fluctuating.] Thus, for large $n$ we
treat the fluctuating quantity $I(x)$ as small, and if we do so we can interpret
$H$ as a time-independent quantum-mechanical Hamiltonian. [The isomonodromic
properties of $H$ when $I(x)$ is not neglected were studied in Ref.~\cite{r6}.]

We conclude that the large-$n$ (semiclassical) behavior of the eigenvalues [that
is, the initial conditions in (\ref{e1})] can be determined by solving the {\it
linear} quantum-mechanical eigenvalue problem $\hat H\psi=E\psi$, where
$\hat H=\half{\hat p}^2-2{\hat x}^3$. To find these eigenvalues we rotate
$\hat H$ into the complex plane \cite{r21} and obtain the well-studied
$\cPT$-symmetric Hamiltonian \cite{r22}
\begin{equation}
\hat H=\half\hat p^2+2i\hat x^3.
\label{e10}
\end{equation}

The large eigenvalues of this Hamiltonian can be found by using the complex WKB
techniques discussed in detail in Ref.~\cite{r22}. For the general class of
$\cPT$-symmetric Hamiltonians $\hat H=\half \hat p^2+g\hat x^2\left(i\hat
x\right)^\epsilon$ $(\epsilon\geq0)$, the WKB approximation to the $n$th
eigenvalue $(n\gg1)$ is given by
\begin{equation}
E_n\sim\frac{1}{2}(2g)^{2/(4+\epsilon)}\left[\frac{\Gamma\left(\frac{3}{2}+\frac
{1}{\epsilon+2}\right)\sqrt{\pi}\,n}{\sin\left(\frac{\pi}{\epsilon+2}\right)
\Gamma\left(1+\frac{1}{\epsilon+2}\right)}\right]^{(2\epsilon+4)/(\epsilon+4)}.
\label{e11}
\end{equation}
Thus, for $H$ in (\ref{e10}) we take $g=2$ and $\epsilon=1$ and obtain
the asymptotic behavior
\begin{equation}
E_n\sim 2\left[\sqrt{3\pi}\Gamma\left(\textstyle{\frac{11}{6}}\right)n/
\Gamma\left(\textstyle{\frac{1}{3}}\right)\right]^{6/5}\quad(n\to\infty).
\label{e12}
\end{equation}

Since $\hat H$ in (\ref{e10}) is time independent, we can evaluate $H$ in
(\ref{e9}) for fixed $y(0)$ and large $y'(0)=b_n$ and obtain the result that
\begin{equation}
b_n\sim\sqrt{2E_n}=B_{\rm I}n^{3/5}\quad(n\to\infty),
\label{e13}
\end{equation}
which verifies (\ref{e6}). We then read off the analytic value of the constant
$B_{\rm I}$:
\begin{equation}
B_{\rm I}=2\left[\sqrt{3\pi}\Gamma\left(\textstyle{\frac{11}{6}}\right)/
\Gamma\left(\textstyle{\frac{1}{3}}\right)\right]^{3/5},
\label{e14}
\end{equation}
which agrees with the numerical result in (\ref{e7}). Also, if we take the
initial slope $y'(0)$ to vanish and take the initial condition $y(0)=c_n$ to be
large, we obtain an analytic expression for $C_{\rm I}$,
\begin{equation}
C_{\rm I}=-\left[\sqrt{3\pi}\Gamma\left(\textstyle{\frac{11}{6}}\right)/
\Gamma\left(\textstyle{\frac{1}{3}}\right)\right]^{2/5},
\label{e15}
\end{equation}
which agrees with the numerical result in (\ref{e8}).

\section{Numerical analysis of the second Painlev\'e transcendent}
\label{s4}
To understand the behavior of solutions to the initial-value problem in
(\ref{e2}) for Painlev\'e II, we follow the procedure used in Sec.~\ref{s2} to
study P-I. An elementary asymptotic analysis shows that as $t\to-\infty$, there
are three possible asymptotic behaviors for solutions $y(t)$. First, $y(t)$ can
oscillate stably about the negative axis. Second, $y(t)$ can approach the curves
$\pm\sqrt{-t/2}$; however, both of these asymptotic behaviors are unstable.

If we numerically integrate (\ref{e2}), we observe that when $t$ becomes large
and negative, a typical solution to the P-II initial-value problem either
oscillates about the negative axis or passes through an infinite sequence of
simple poles. However, it is also possible to find special eigenfunction
solutions that pass through only a finite number of poles and then approach
either the positive or the negative branches of the square-root curves. These
eigenfunctions obey the boundary conditions $y(0)=0$ and $y'(0)=\pm b$. [Note
that P-II is symmetric under $y\to-y$, so there are two sets of eigenfunctions,
one for each sign of $y'(0)$.] We study these eigenfunctions numerically in
Subsec.~\ref{ss4a}. The P-II equation is particularly interesting because as
$t\to+\infty$, the behavior $y\to0$ becomes unstable. Thus, it is possible to
have new kinds of eigenfunctions for positive $t$ as well. We seek
eigenfunctions that satisfy $y'(0)=0$ and $y(0)=c$ and examine the positive-$c$
eigenfunctions numerically in Subsec.~\ref{ss4b}.

\subsection{Initial-slope eigenvalues for Painlev\'e I}
\label{ss4a}

\begin{figure}[h!]
\null\vspace{-9mm}
\begin{center}
\includegraphics[trim=21mm 50mm 18mm 50mm,clip=true,scale=0.45]{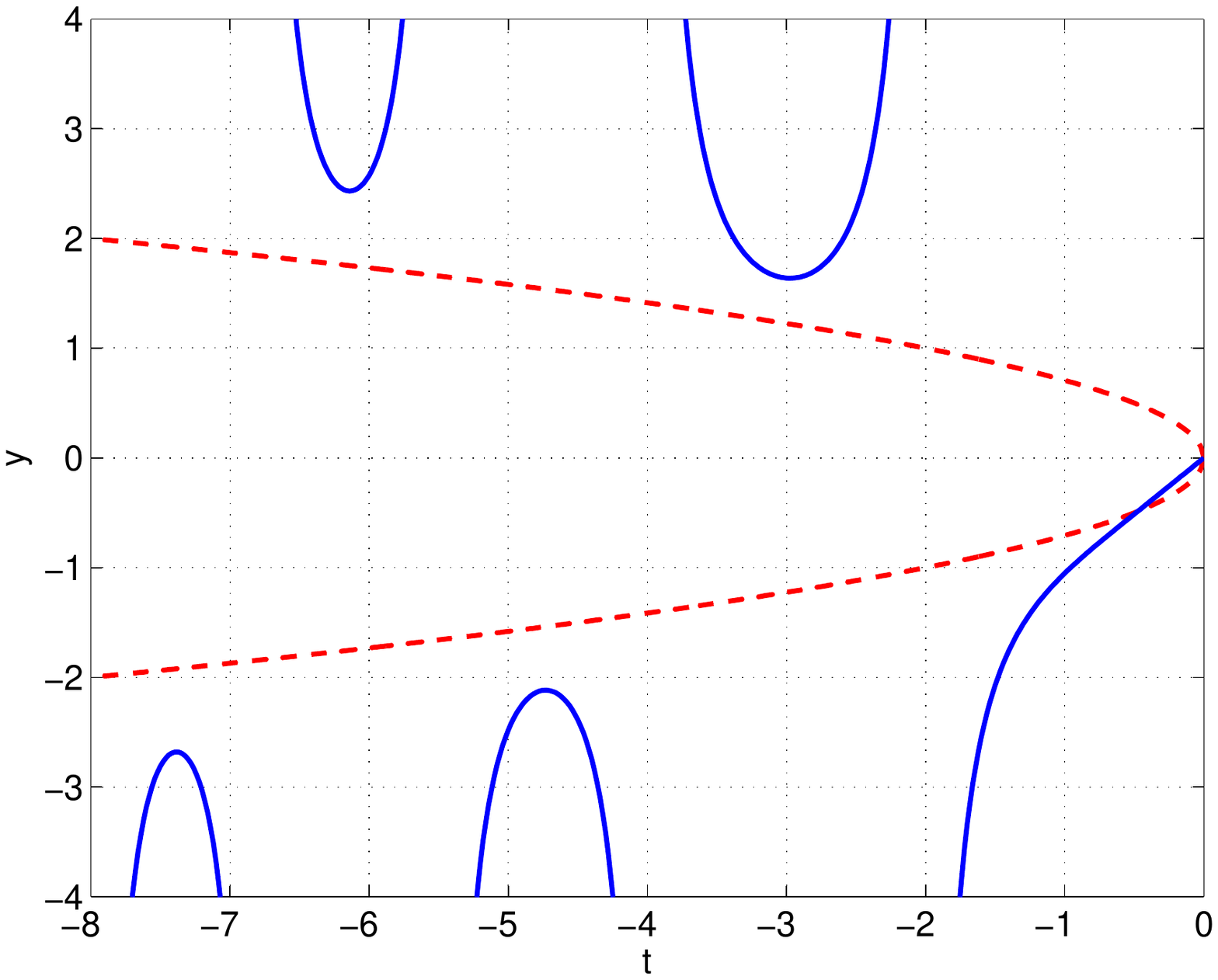}
% trim=left bottom right top
\includegraphics[trim=18mm 50mm 21mm 50mm,clip=true,scale=0.45]{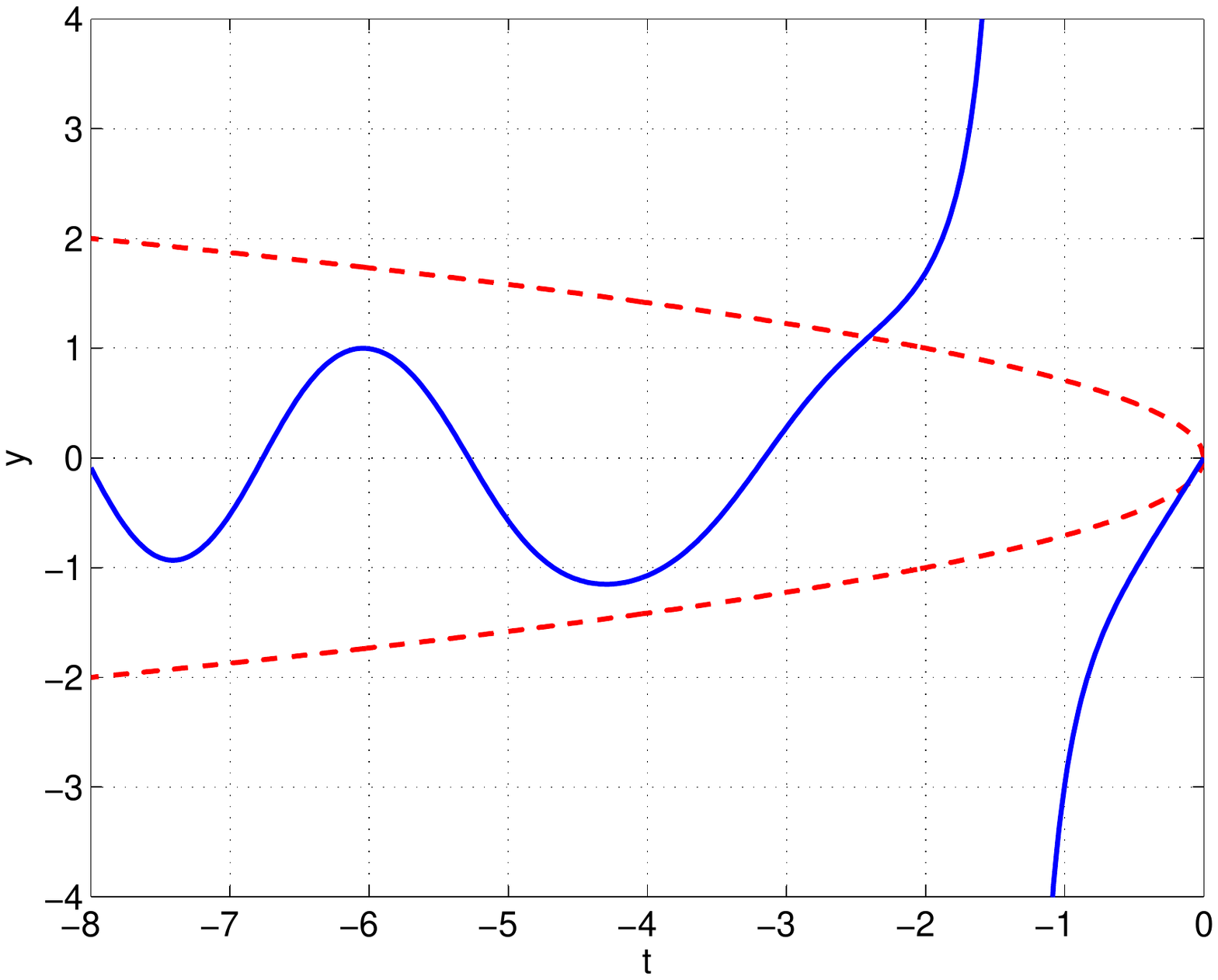}
\hspace{1.5cm}
\end{center}
\null\vspace{-21mm}
\caption{Typical behavior of solutions to the second Painlev\'e transcendent for
the initial conditions $y(0)=0$ and $b=y'(0)$. In the left panel $b=
1.028605106$, which lies between the eigenvalues $b_1=0.5950825526$ and $b_2=
1.528605106$. In the right panel $b=2.028605106$, which lies between the
eigenvalues $b_2=1.528605106$ and $b_3=2.155132869$. In the left panel the
solution $y(t)$ has an infinite sequence of simple poles and in the right panel
the solution oscillates stably about $-\sqrt{t/6}$. The dashed curves are the
functions $\pm\sqrt{-t/2}$.}
\label{F8}
\end{figure}

\begin{figure}[h!]
\null\vspace{-9mm}
\begin{center}
\includegraphics[trim=21mm 50mm 18mm 50mm,clip=true,scale=0.45]{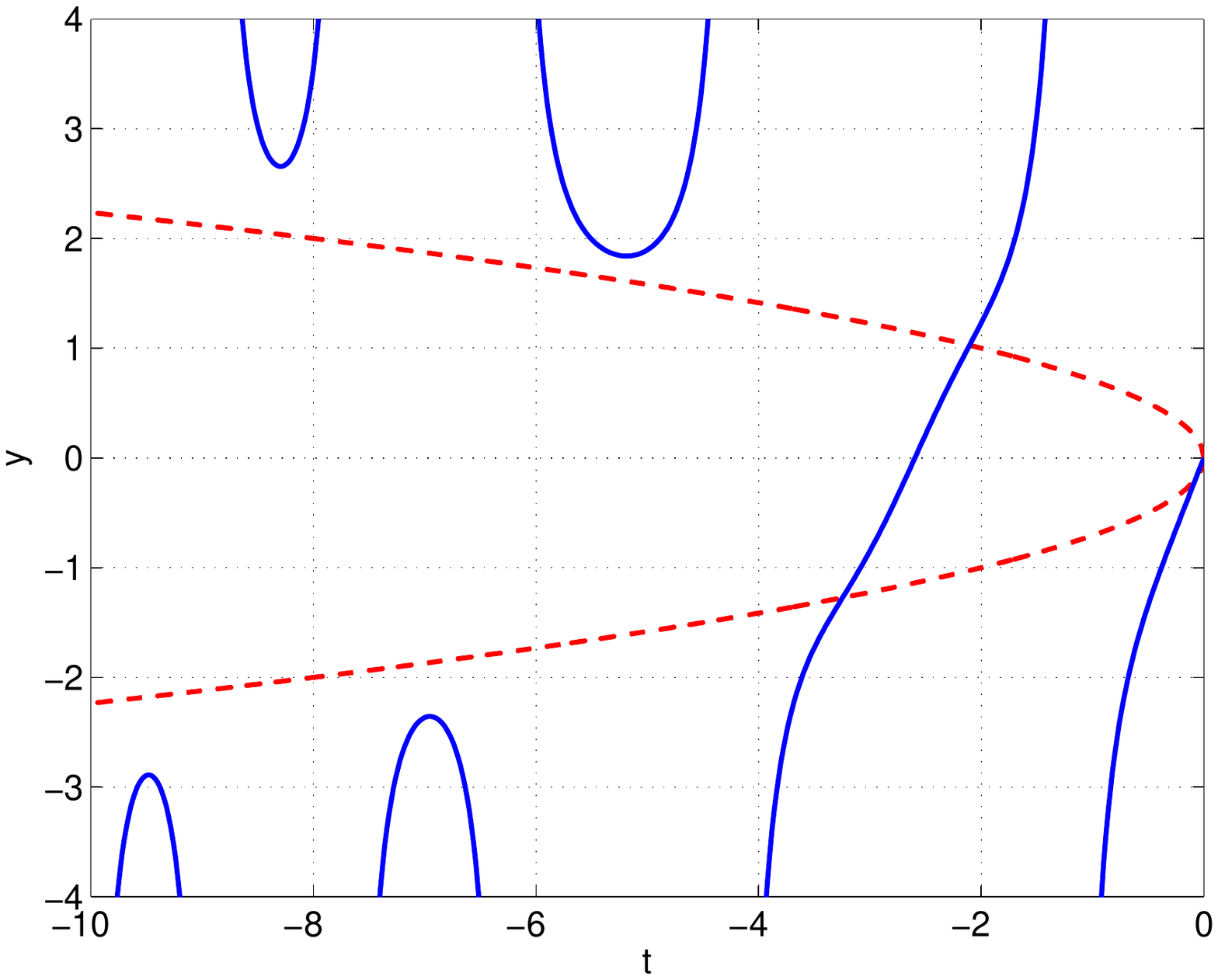}
% trim=left bottom right top
\includegraphics[trim=18mm 50mm 21mm 50mm,clip=true,scale=0.45]{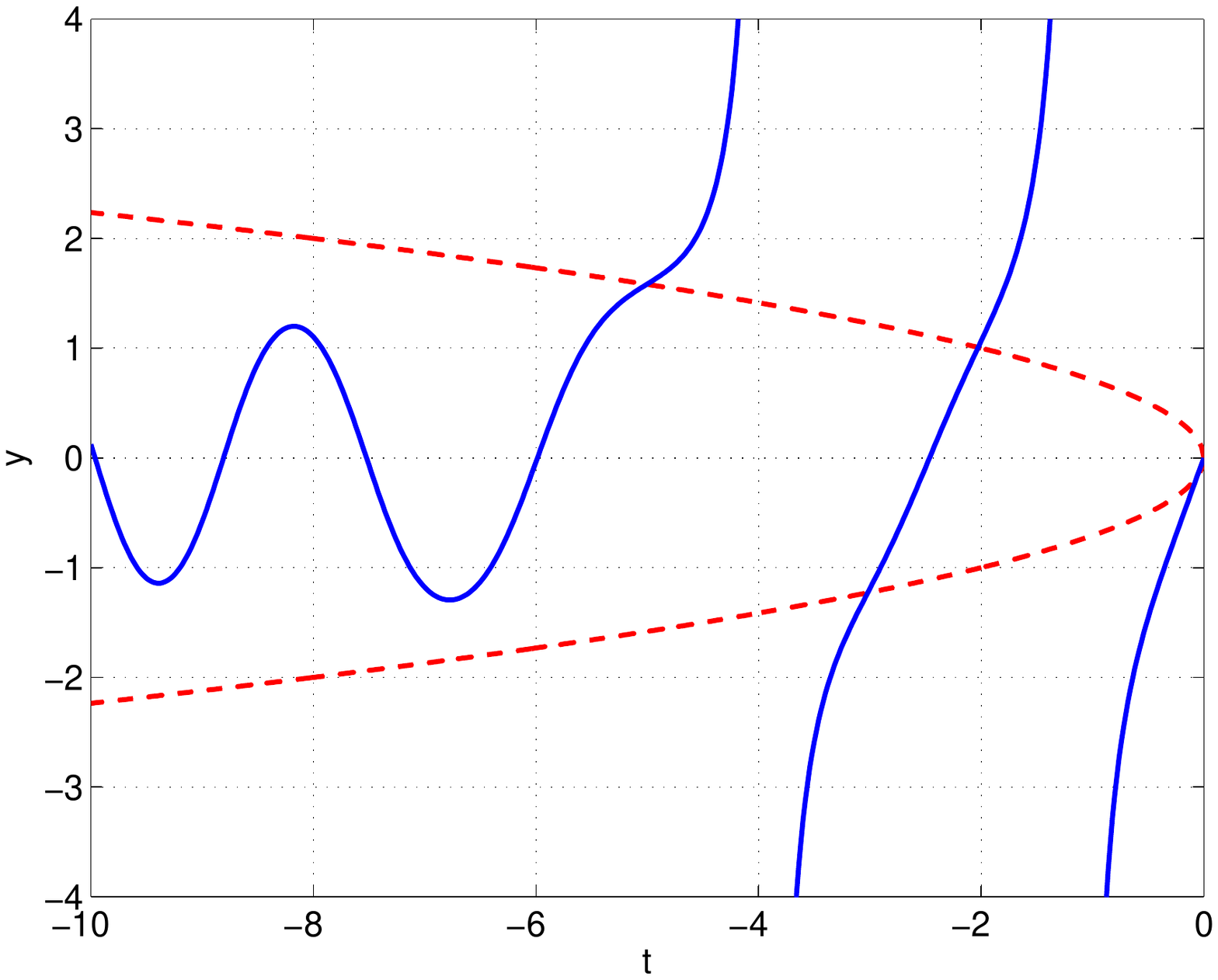}
\hspace{1.5cm}
\end{center}
\null\vspace{-21mm}
\caption{Solutions to the P-II equation (\ref{e2}) for $y(0)=0$ and $b=y'(0)$.
Left panel: $b=2.600745985$, which lies between the eigenvalues $b_3=
2.155132869$ and $b_4=2.700745985$. Right panel: $b=2.800745985$, which
lies between the eigenvalues $b_4=2.700745985$ and $b_5=3.195127590$.}
\label{F9}
\end{figure}

\begin{figure}[t!]
\null\vspace{-9mm}
\begin{center}
\includegraphics[trim=21mm 50mm 18mm 50mm,clip=true,scale=0.45]{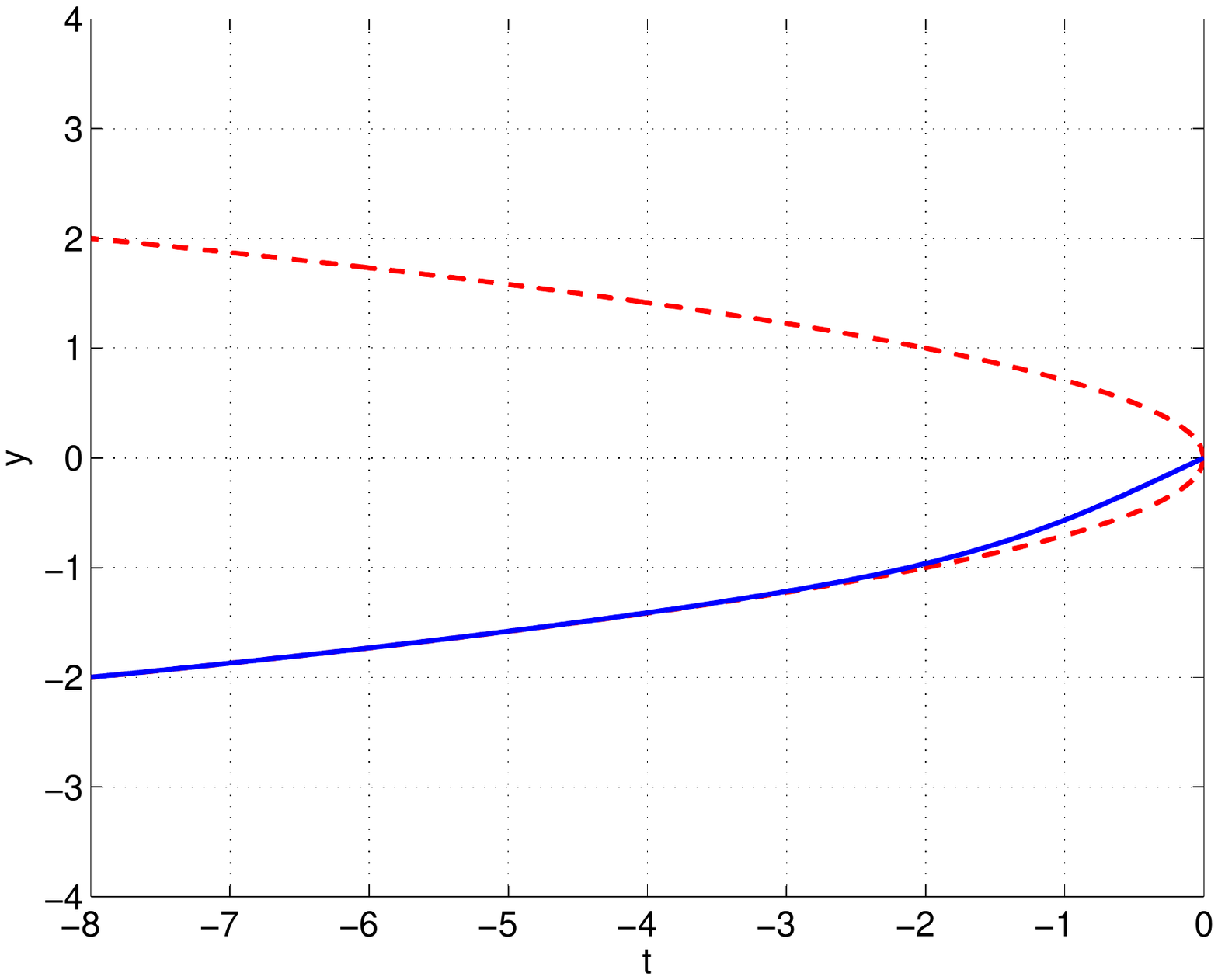}
% trim=left bottom right top
\includegraphics[trim=18mm 50mm 21mm 50mm,clip=true,scale=0.45]{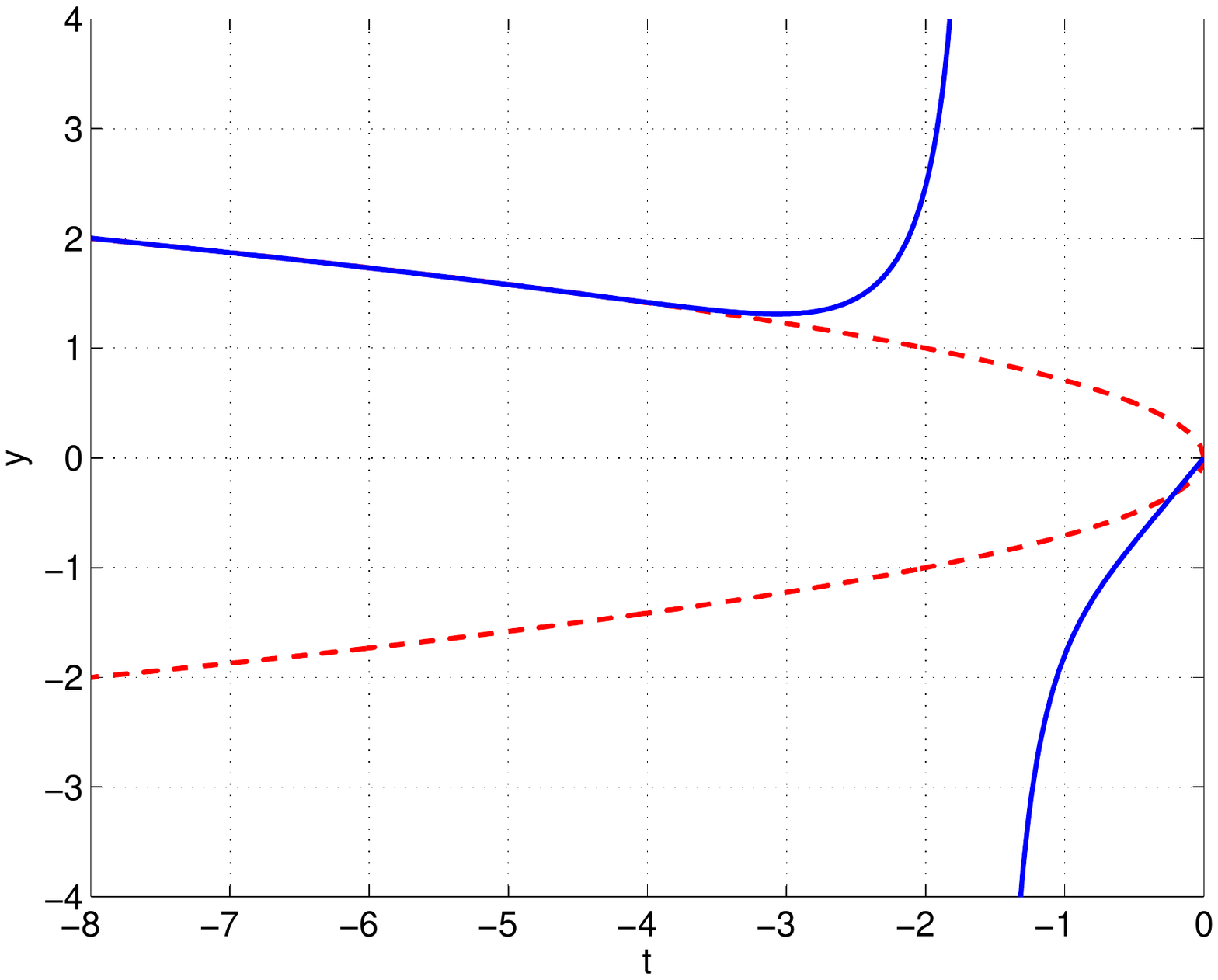}
\hspace{1.5cm}
\end{center}
\null\vspace{-21mm}
\caption{First two separatrix solutions (eigenfunctions) of Painlev\'e II with
initial condition $y(0)=0$. Left panel: $y'(0)=b_1=0.5950825526$; right panel:
$y'(0)=b_2=1.528605106$. The dashed curves are $\pm\sqrt{-t/2}$.}
\label{F10}
\end{figure}

\begin{figure}[h!]
\null\vspace{-9mm}
\begin{center}
\includegraphics[trim=21mm 50mm 18mm 50mm,clip=true,scale=0.45]{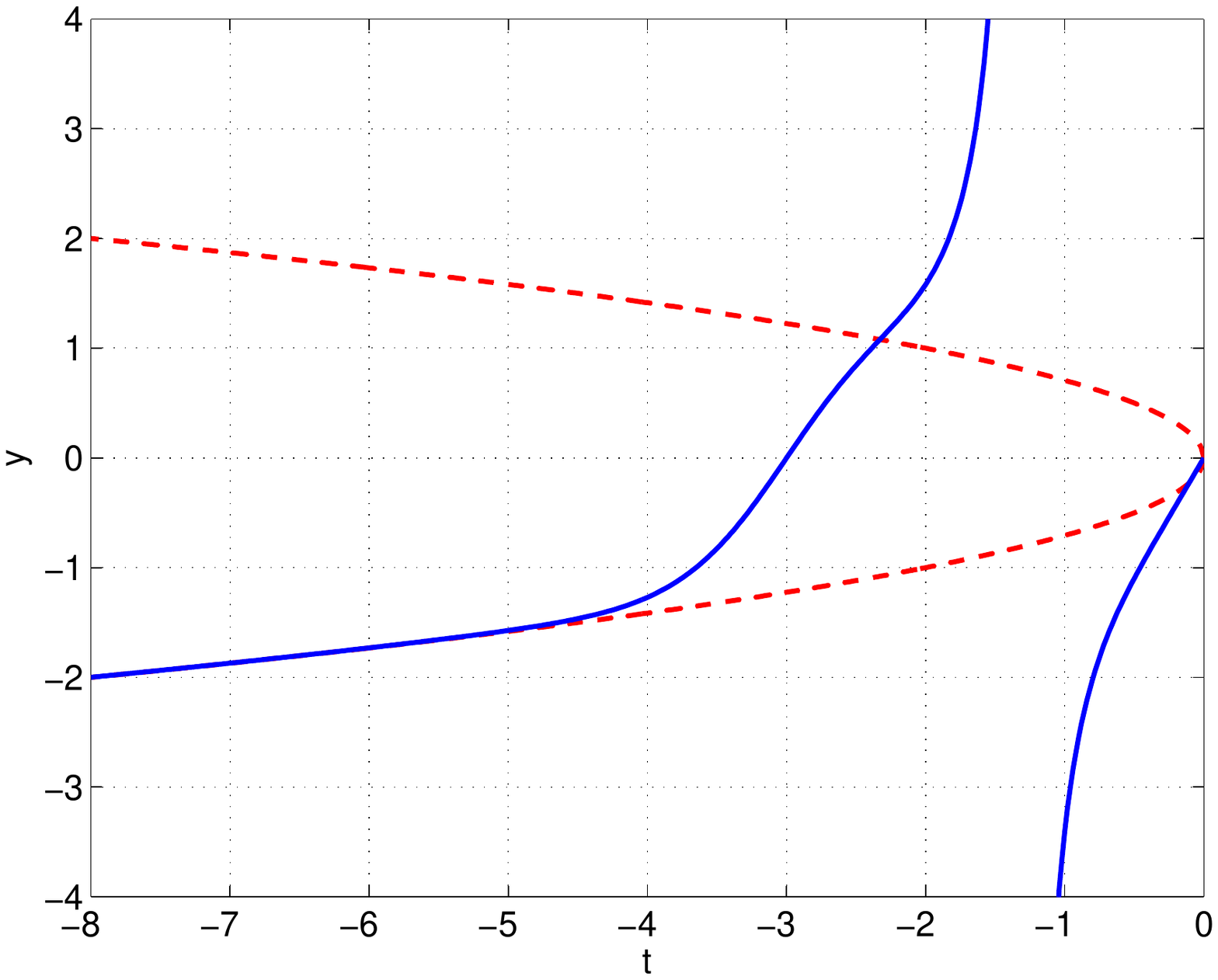}
% trim=left bottom right top
\includegraphics[trim=18mm 50mm 21mm 50mm,clip=true,scale=0.45]{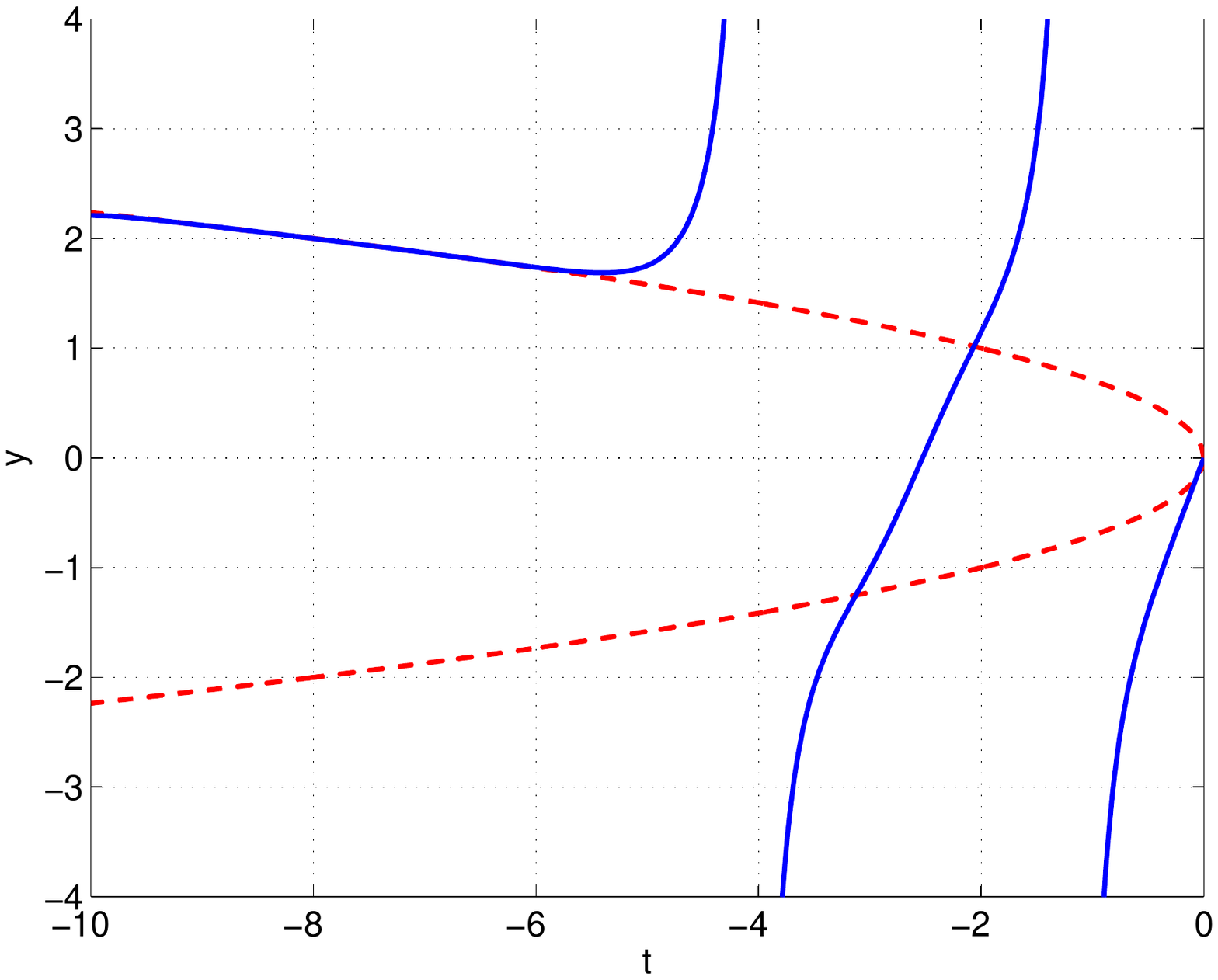}
\hspace{1.5cm}
\end{center}
\null\vspace{-21mm}
\caption{Third and fourth eigenfunctions of Painlev\'e II with initial condition
$y(0)=0$. Left panel: $y'(0)=b_3=2.155132869$; right panel: $y'(0)=b_4=
2.700745985$.}
\label{F11}
\end{figure}

\begin{figure}[t!]
\null\vspace{-9mm}
\begin{center}
\includegraphics[trim=21mm 50mm 18mm 50mm,clip=true,scale=0.45]{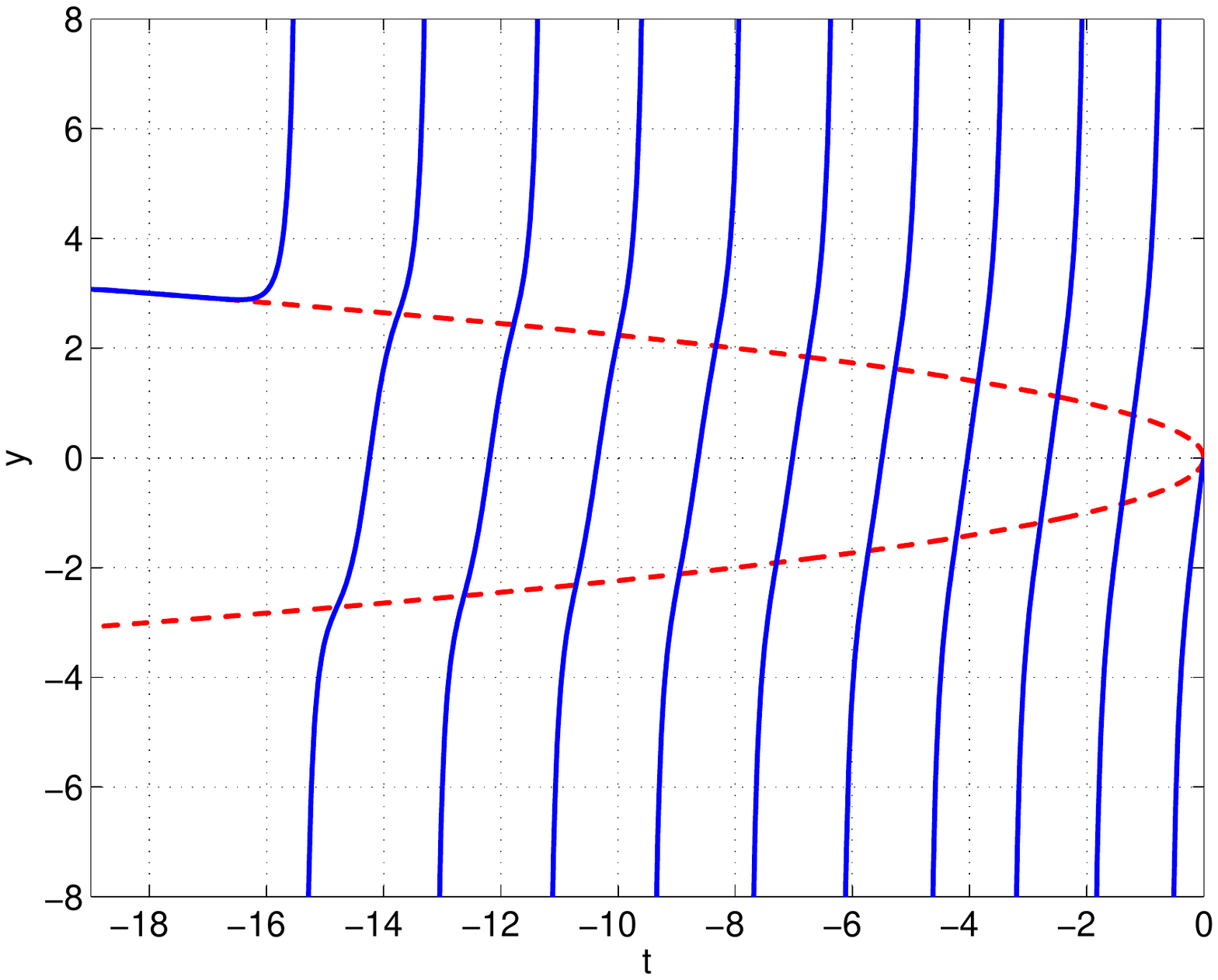}
% trim=left bottom right top
\includegraphics[trim=18mm 50mm 21mm 50mm,clip=true,scale=0.45]{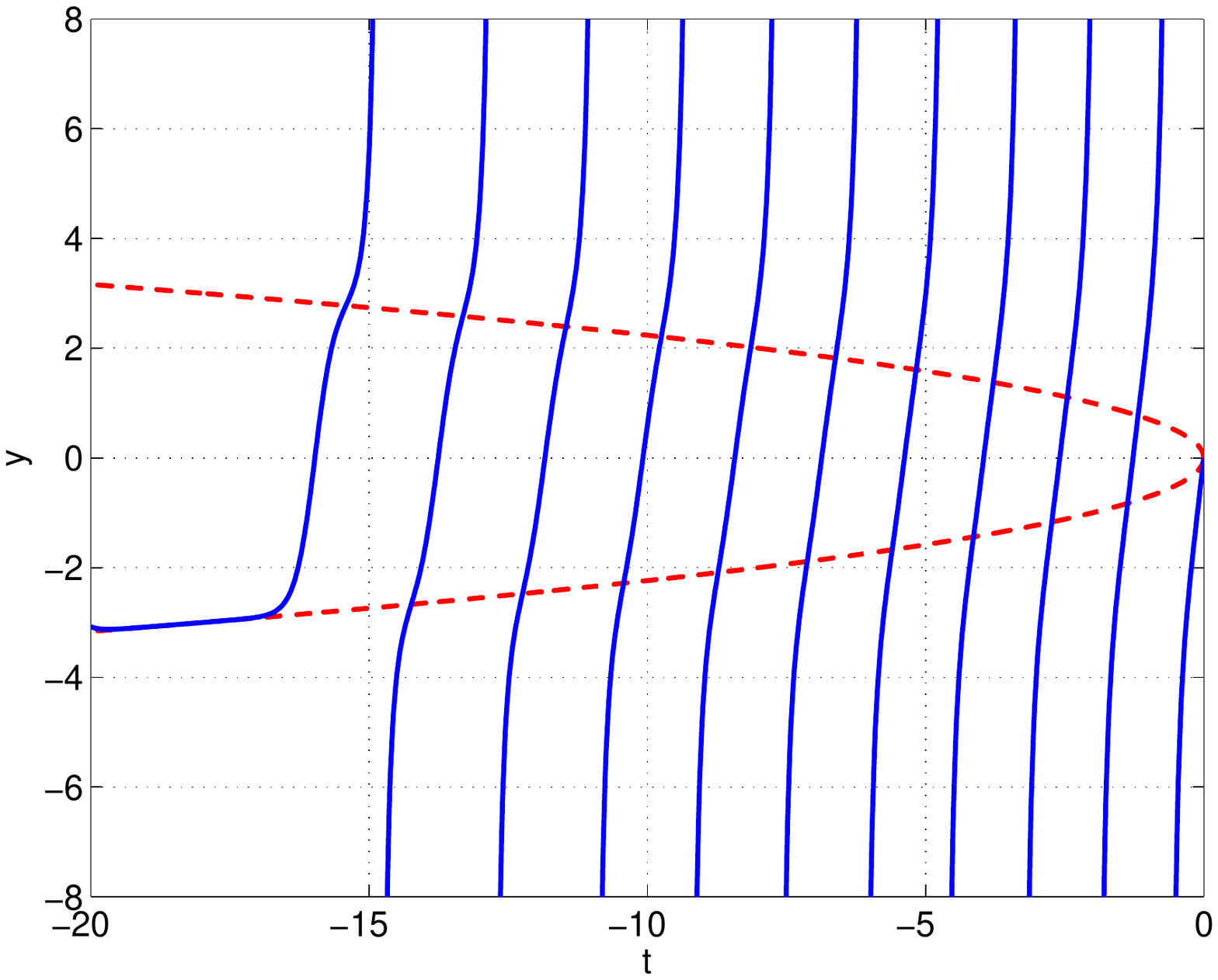}
\hspace{1.5cm}
\end{center}
\null\vspace{-21mm}
\caption{The twentieth and twenty-first eigenfunctions of Painlev\'e II with
initial condition $y(0)=0$. Left panel: $y'(0)=b_{20}=8.499476190$; right panel:
$y'(0)=b_{21}=8.787666814$.}
\label{F12}
\end{figure}

Similar to what we found in Sec.~\ref{s2}, if we choose $y(0)=0$, there are
critical values $y'(0)=b_n$ at which the solutions $y(t)$ change their
character. In Figs.~\ref{F8} and \ref{F9} we plot the solutions to the P-II
equation for the initial condition $y(0)=0$ and $y'(0)=b$ for $b_1<b<b_2$, $b_2<
b<b_3$, $b_3<b<b_4$, and $b_4<b<b_5$. Note that in these figures the character
of the solution alternates between having an infinite sequence of simple poles
and oscillating stably about $y(t)=0$. However, when $y'(0)=b$ is at a critical
value (eigenvalue) $b_n$, the solution $y(t)$ passes through a {\it finite}
number $[n/2]$ of simple poles and then approaches either $+\sqrt{-t/2}$ or
$-\sqrt{-t/2}$. These eigenfunctions (separatrices) are plotted in
Figs.~\ref{F10}, \ref{F11}, and \ref{F12} for $n=(1,2)$, $(3,4)$, and $(20,21)$.

Note that the eigenfunctions in Figs.~\ref{F10}, \ref{F11}, and \ref{F12}
alternate between approaching the upper-unstable branch $+\sqrt{-t/2}$ or the
lower-unstable branch $-\sqrt{-t/2}$, and thus there are actually two sequences
of eigenvalues, one for even $n$ and one for odd $n$. Using Richardson
extrapolation, we find that the sequences of eigenvalues $b_{2n}$ and
$b_{2n+1}$ have the same asymptotic behavior
\begin{equation}
b_{2n}\sim b_{2n+1}\sim B_{\rm II}n^{2/3}\quad(n\to\infty).
\label{e16}
\end{equation}
Our numerical calculations give
\begin{equation}
B_{\rm II}=1.862412{\underbar 8}.
\label{e17}
\end{equation}
The numerical data for P-II are slightly more noisy than those for P-I, and
fourth-order Richardson extrapolation only gives the underlined eighth digit as
$8\pm2$.

\subsection{Initial-value eigenvalues for Painlev\'e II}
\label{ss4b}
\begin{figure}[h!]
\null\vspace{-9mm}
\begin{center}
\includegraphics[trim=21mm 50mm 18mm 50mm,clip=true,scale=0.45]{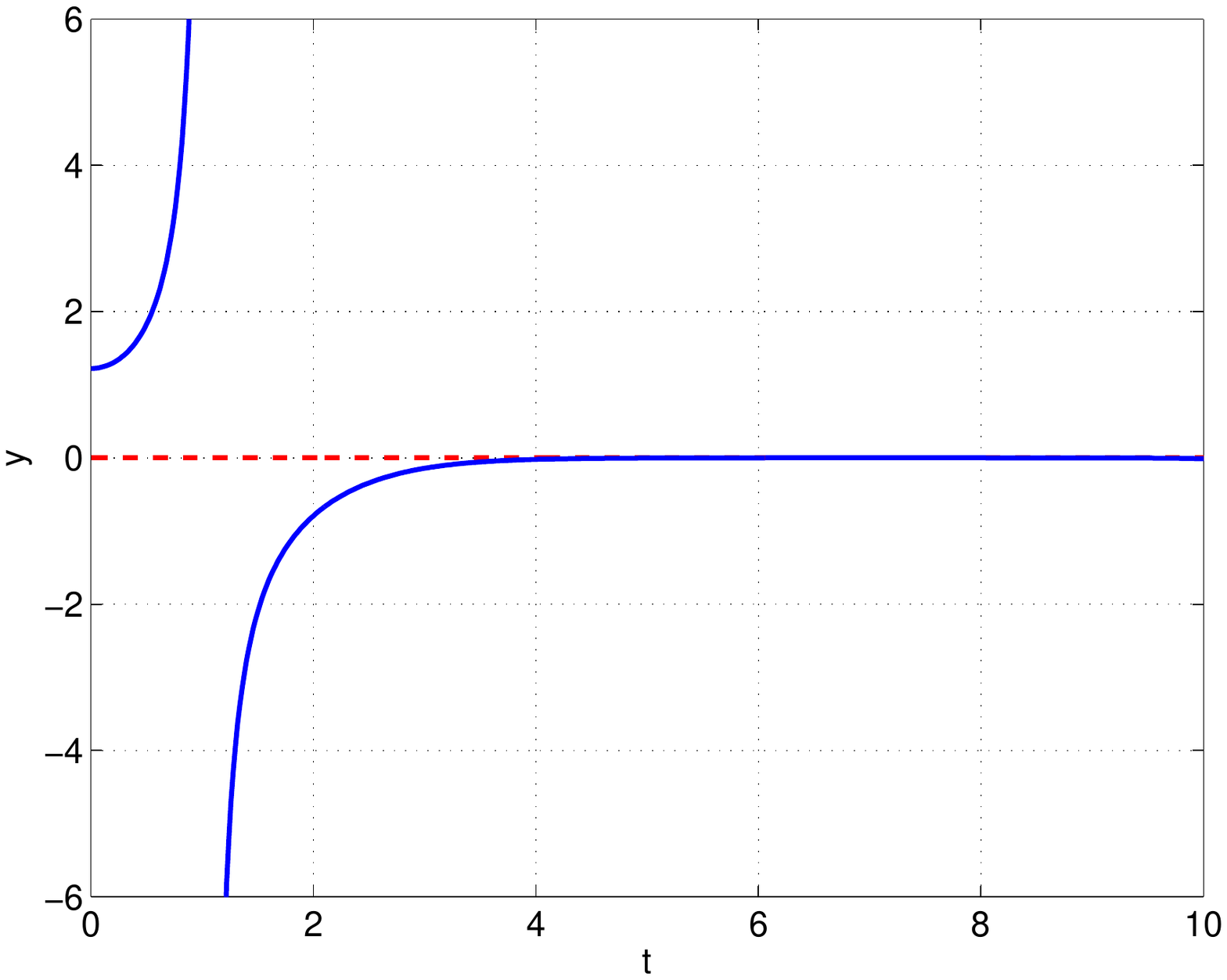}
% trim=left bottom right top
\includegraphics[trim=18mm 50mm 21mm 50mm,clip=true,scale=0.45]{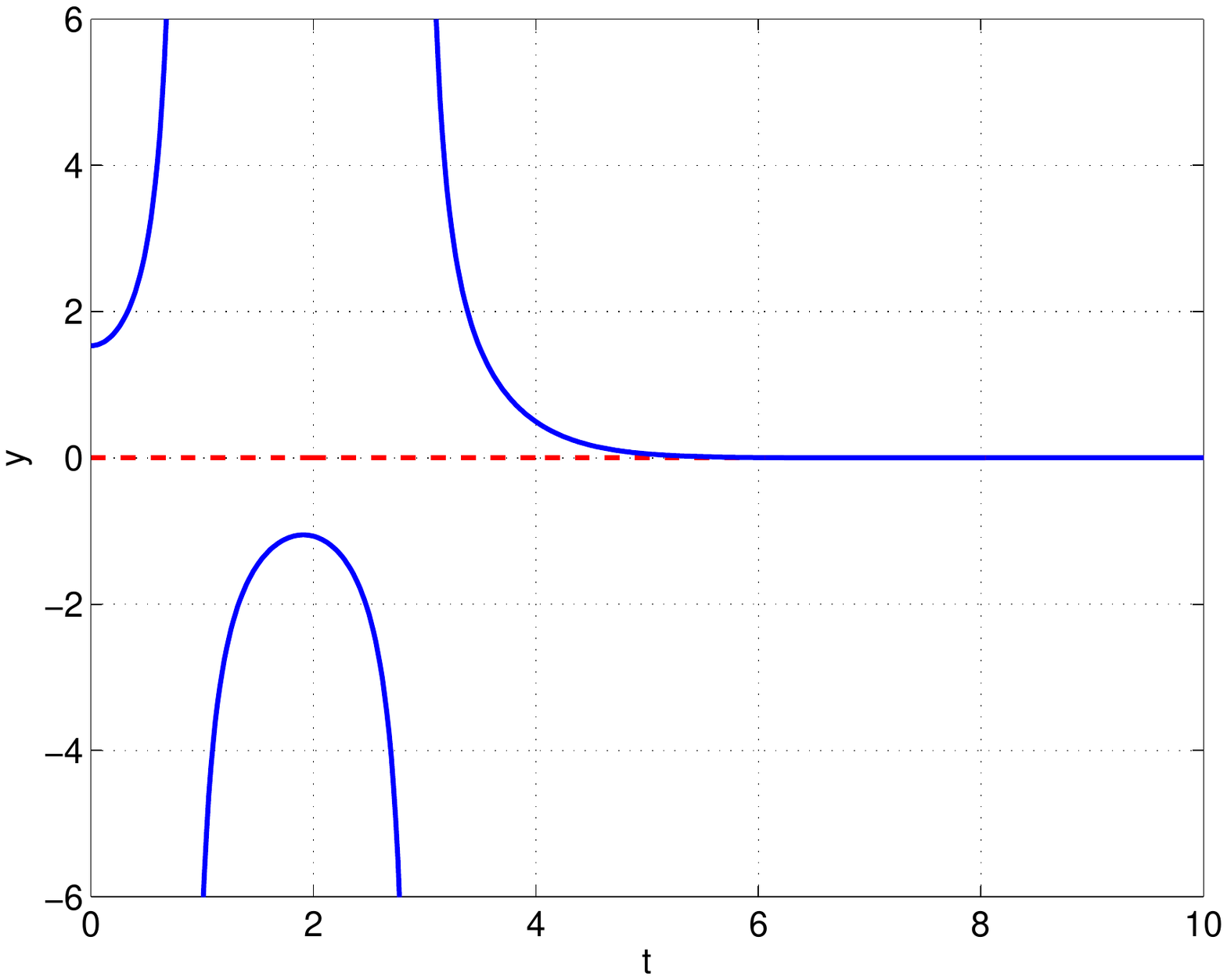}
\hspace{1.5cm}
\end{center}
\null\vspace{-21mm}
\caption{First two separatrix solutions (eigenfunctions) of Painlev\'e II with
fixed initial slope $y'(0)=0$. Left panel: $y(0)=c_1=1.222873339$; right
panel: $y(0)=c_2=1.533883935$.}
\label{F13}
\end{figure}

\begin{figure}[h!]
\null\vspace{-9mm}
\begin{center}
\includegraphics[trim=21mm 50mm 18mm 50mm,clip=true,scale=0.45]{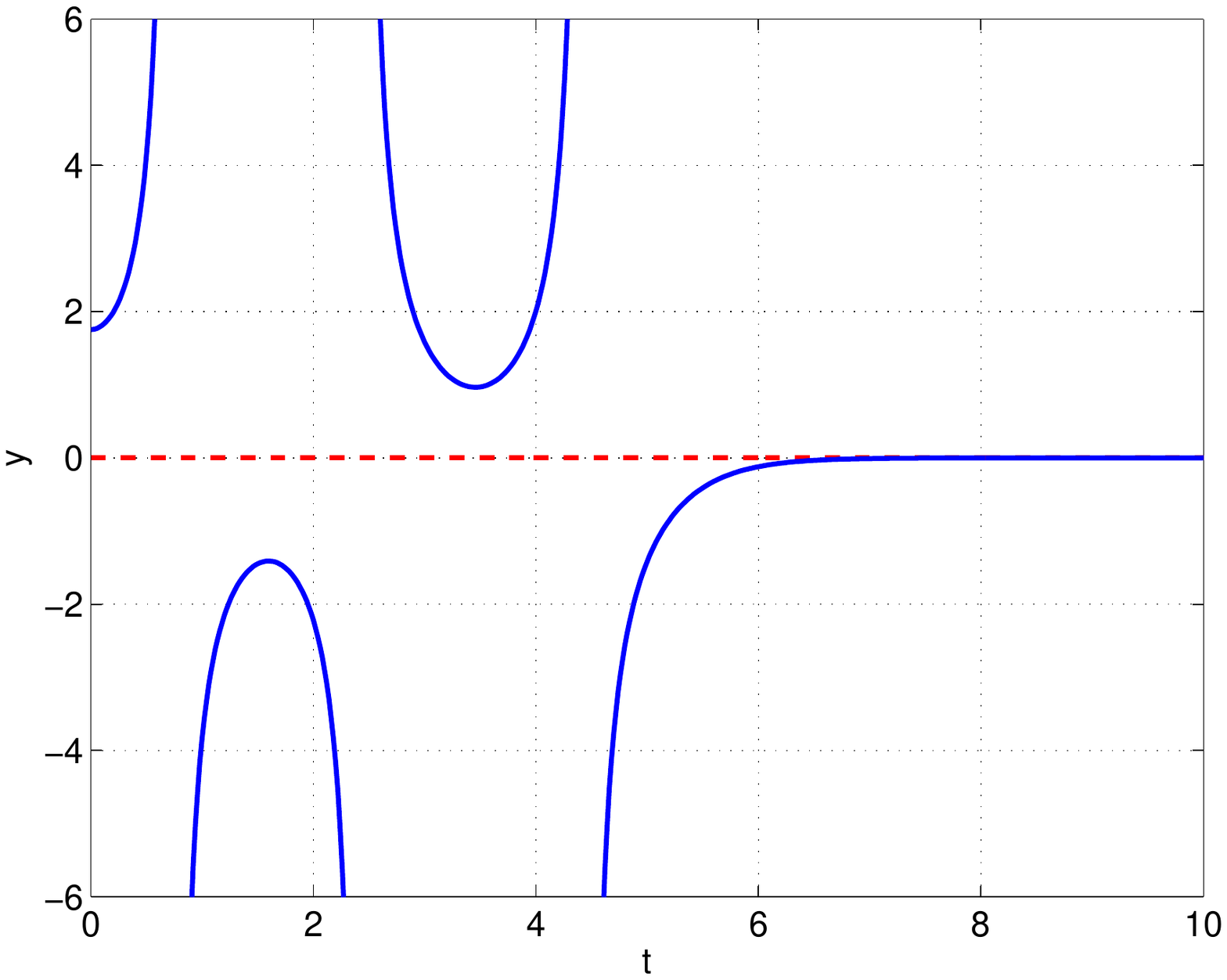}
% trim=left bottom right top
\includegraphics[trim=18mm 50mm 21mm 50mm,clip=true,scale=0.45]{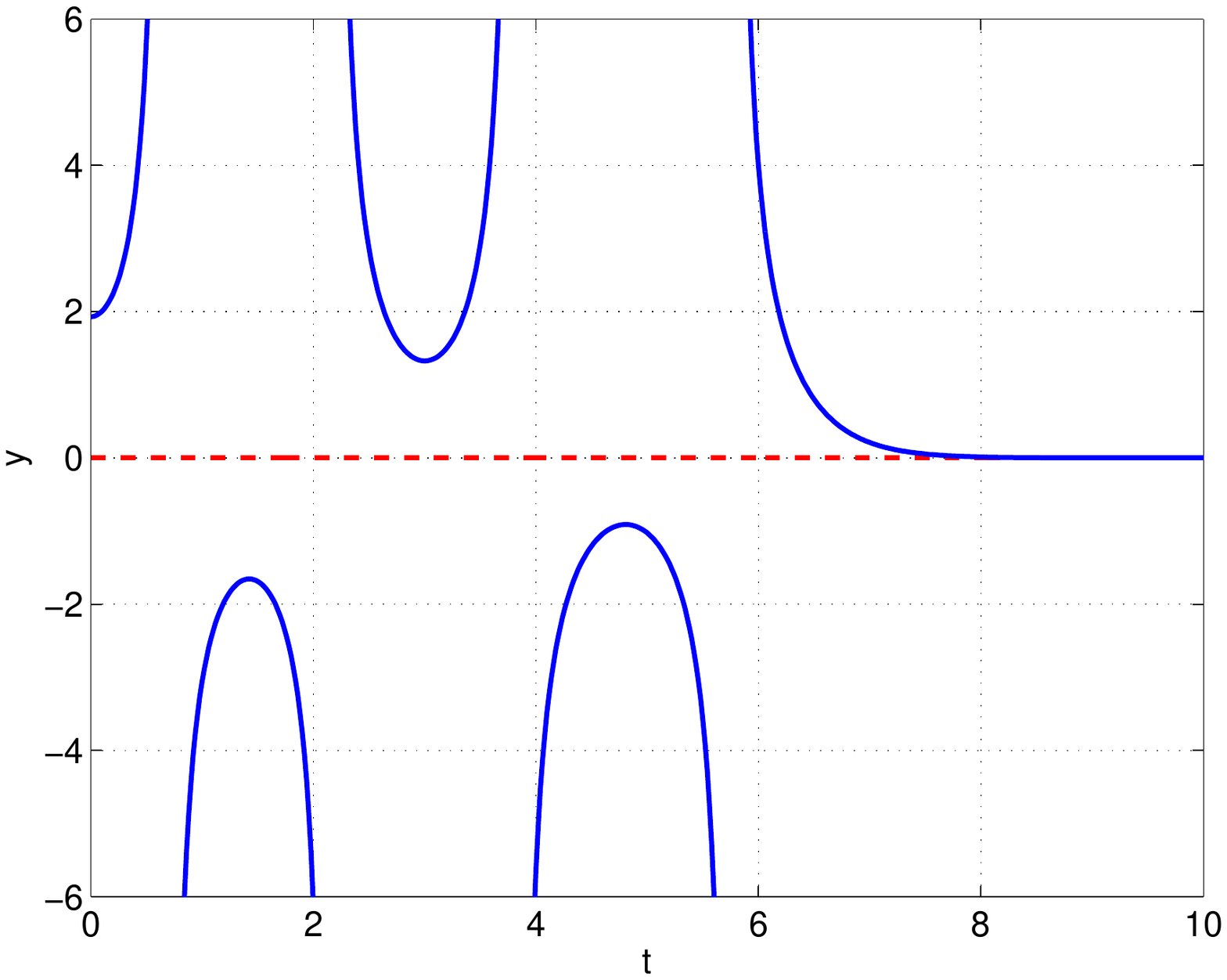}
\hspace{1.5cm}
\end{center}
\null\vspace{-21mm}
\caption{Third and fourth eigenfunctions of Painlev\'e II with initial slope $y'(
0)=0$. Left panel: $y(0)=c_3=1.754537281$; right panel: $y(0)=c_4=
1.93061783$.}
\label{F14}
\end{figure}

\begin{figure}[h!]
\null\vspace{-9mm}
\begin{center}
\includegraphics[trim=21mm 50mm 18mm 50mm,clip=true,scale=0.45]{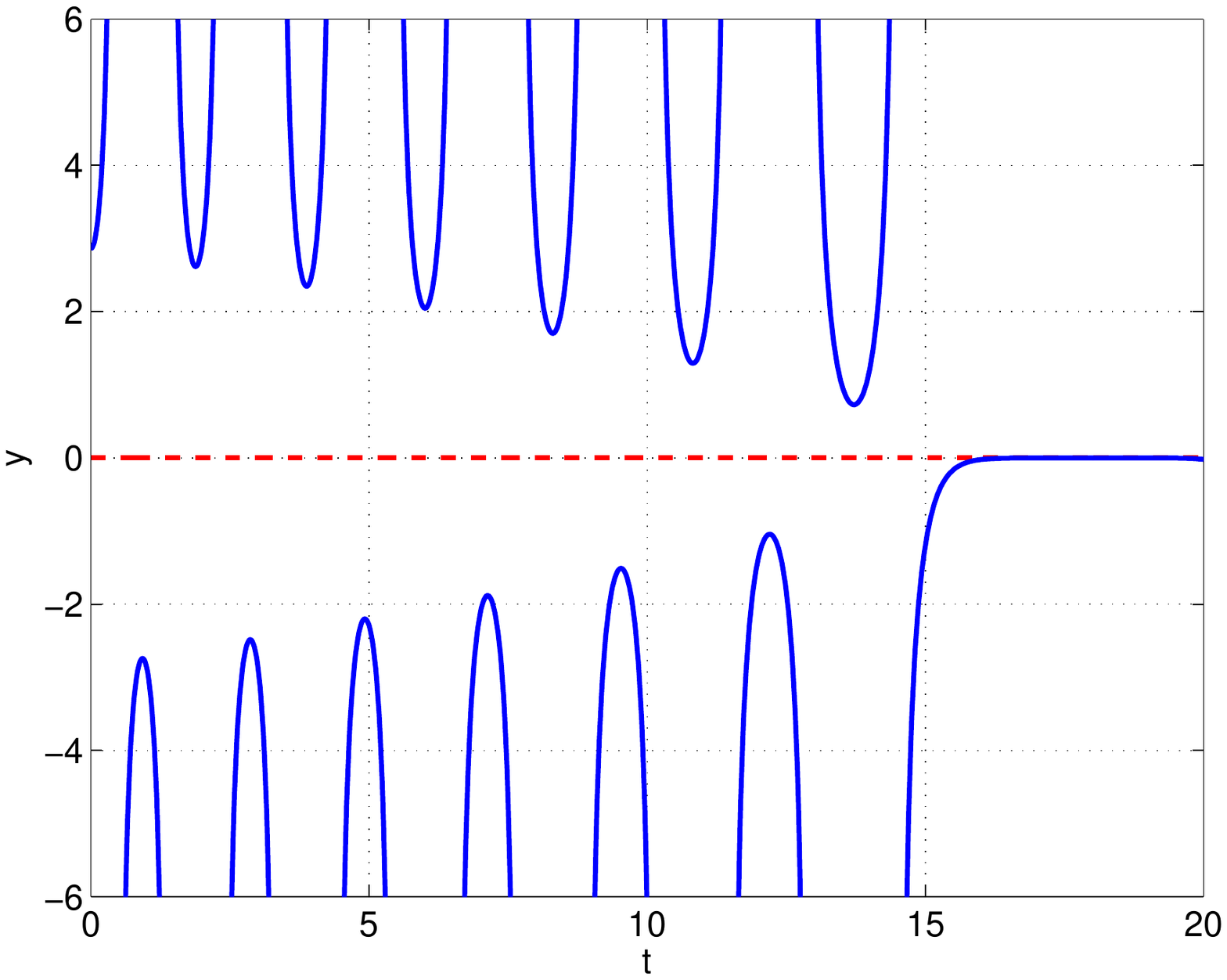}
% trim=left bottom right top
\includegraphics[trim=18mm 50mm 21mm 50mm,clip=true,scale=0.45]{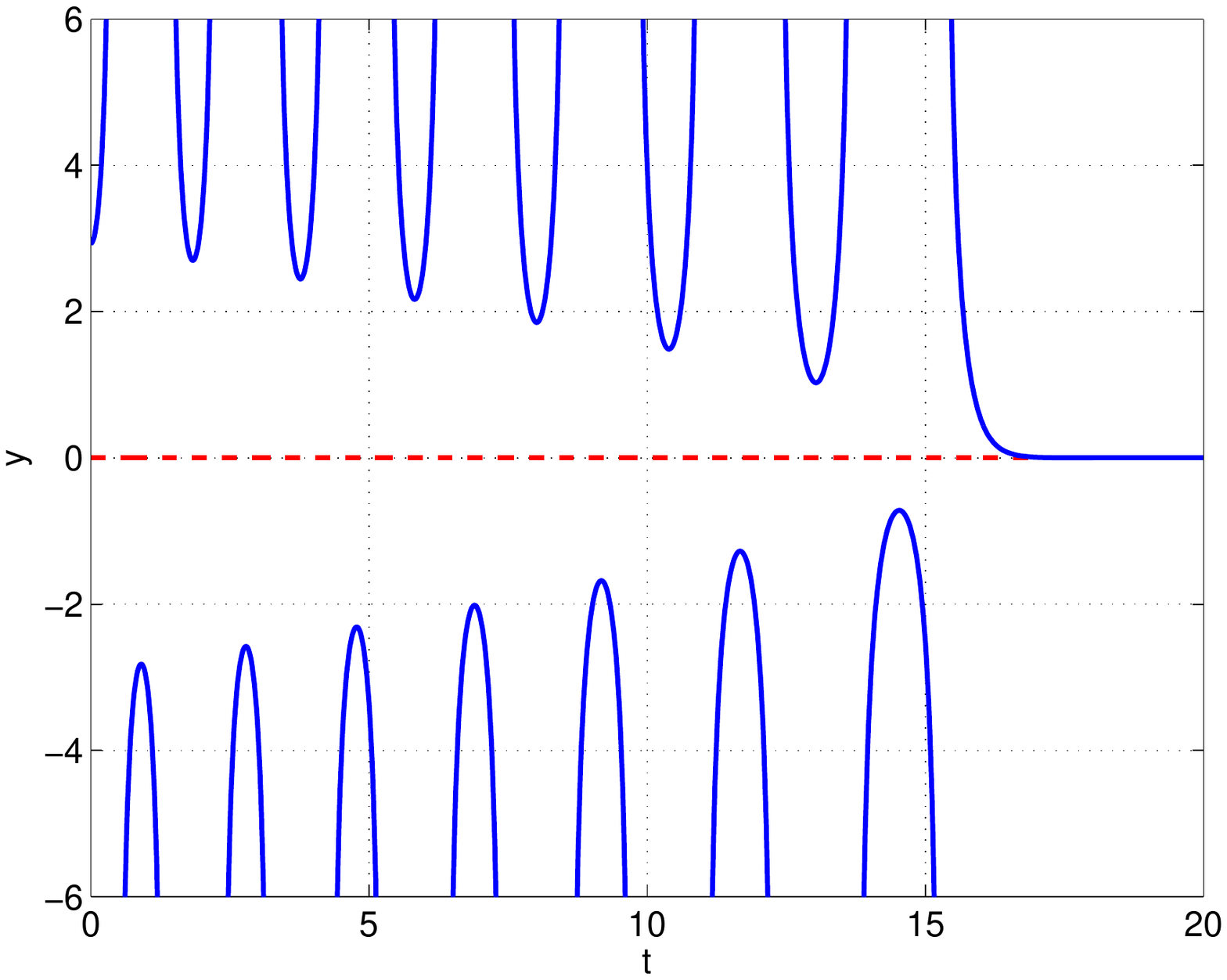}
\hspace{1.5cm}
\end{center}
\null\vspace{-21mm}
\caption{Thirteenth and fourteenth separatrix solutions (eigenfunctions) of
Painlev\'e II with fixed initial slope $y'(0)=0$. Left panel: $y(0)=c_1=
2.858869051$; right panel: $y(0)=c_2=2.9303576515$.}
\label{F15}
\end{figure}

Next, we plot the positive-$t$ solutions to P-II for vanishing initial slope and
positive initial condition for $t\geq0$. As $t\to\infty$, the $n$th
eigenfunction passes through $n$ simple poles before it approaches zero
monotonically. In Figs.~\ref{F13}, \ref{F14}, and \ref{F15} we plot the six
eigenfunctions corresponding to $n=(1,2)$, $(3,4)$, and $(13,14)$. (Because of
the symmetry of P-II, for every positive eigenvalue there is a corresponding
negative eigenvalue. We do not plot the negative-eigenvalue solutions.)

Using fourth-order Richardson we determine that for large $n$, $c_n\sim
C_{\rm II}n^{1/3}$, where
\begin{equation}
C_{\rm II}=1.2158116{\underbar 5}.
\label{e18}
\end{equation}
The last digit $5$ has an uncertainty of $\pm1$.

\section{Asymptotic calculation of $B_{\rm II}$ and $C_{\rm II}$}
\label{s5}
To obtain analytic expressions for $B_{\rm II}$ in (\ref{e17}) and $C_{\rm II}$
in (\ref{e18}), we follow the same procedure as in Sec.~\ref{s3} for P-I. We
multiply the P-II differential equation in (\ref{e2}) by $y'(t)$ and
integrate from $t=0$ to $t=x$, where $x$ is in the turning-point region
which the simple poles stop. The result is
\begin{equation}
H\equiv\half[y'(x)]^2-\half[y(x)]^4=\half[y'(0)]^2-\half[y(0)]^4+I(x),
\label{e19}
\end{equation}
where $I(x)=\int_0^x dt\,ty(t)y'(t)$. The path of integration is the same as
that used to calculate P-II numerically in Sec.~\ref{s4}; it follows the
negative-real axis until it gets near a simple pole, at which point it makes a
semicircular detour in the complex-$t$ plane to avoid the pole. Again, as in
Sec.~\ref{s3}, we argue that along this path the integrand of $I(x)$ is
oscillatory and because of cancellations we may neglect $I(x)$ when $n$ is
large.

We treat $H$ as the $\cPT$-symmetric quantum-mechanical Hamiltonian
\begin{equation}
\hat H=\half\hat p^2-\half\hat x^4
\label{e20}
\end{equation}
and we use (\ref{e11}) with $g=1/2$ and $\epsilon=2$ to obtain the formula
\begin{equation}
E_n\sim\half\left[3n\sqrt{2\pi}\Gamma\left(\textstyle{\frac{3}{4}}\right)/\Gamma
\left(\textstyle{\frac{1}{4}}\right)\right]^{4/3}
\label{e21}
\end{equation}
for the large eigenvalues of $\hat H$. Finally, we calculate the eigenvalues
$b_n$ by using
\begin{equation}
\sqrt{2E_n}\sim\left[3n\sqrt{2\pi}\Gamma\left(\textstyle{\frac{3}{4}}\right)/
\Gamma\left(\textstyle{\frac{1}{4}}\right)\right]^{2/3}\quad(n\to\infty).
\label{e22}
\end{equation}
This result allows us to identify the value of $B_{\rm II}$ in (\ref{e17}) as
\begin{equation}
B_{\rm II}=\left[3\sqrt{2\pi}\Gamma\left(\textstyle{\frac{3}{4}}\right)/\Gamma
\left(\textstyle{\frac{1}{4}}\right)\right]^{2/3}.
\label{e23}
\end{equation}
This result agrees with the numerical determination in (\ref{e17}).

To calculate $C_{II}$ we observe from Figs.~\ref{F13}-\ref{F15} that the initial
value $y(0)$ is positive. However, if we neglext $I(x)$ and assume a vanishing
initial slope, we see that the right side of (\ref{e19}) negative. Thus, as
we did for the cubic Hamiltonian $\half\hat p^2-2\hat x^3$, we perform a complex
rotation of the coupling constant to convert the quartic Hamiltonian to the
form
\begin{equation}
\hat H=\half\hat p^2+\half\hat x^4.
\label{e24}
\end{equation}
This is the conventional Hermitian quartic-anharmonic-oscillator Hamiltonian,
and does not belong to the class of $\cPT$-symmetric Hamiltonians $\hat H=\half
\hat p^2+g\hat x^2(i\hat x)^\epsilon$. A WKB calculation gives the
large-eigenvalue approximation
\begin{equation}
E_n\sim\left[3n\sqrt{\pi}\Gamma\left(\textstyle{\frac{3}{4}}\right)/
\Gamma\left(\textstyle{\frac{1}{4}}\right)\right]^{4/3}\quad(n\to\infty).
\label{e25}
\end{equation}
Thus, we read off the value of $C_{II}$:
\begin{equation}
C_{II}=\left[3\sqrt{\pi}\Gamma\left(\textstyle{\frac{3}{4}}\right)/
\Gamma\left(\textstyle{\frac{1}{4}}\right)\right]^{1/3},
\label{e26}
\end{equation}
which agrees exactly with the numerical result in (\ref{e18}).

\section{Brief concluding remarks}
\label{s6}

In this paper we have shown that the first two Painlev\'e equations, P-I and
P-II, exhibit instabilities that are associated with separatrix solutions. The 
initial conditions that give rise to these separatrix solutions are eigenvalues.
We have calculated the semiclassical (large-eigenvalue) behavior of the
eigenvalues in two ways, first by using numerical techniques and then by using
asymptotic methods to reduce the initial-value problems for the nonlinear P-I
and P-II equations to linear eigenvalue problems associated with the
time-independent Schr\"odinger equation. The agreement between these two
approaches is exact.
 
The obvious continuation of this work is to examine the next four Painlev\'e
equations, P-III --- P-VI, to see if there are instabilities, separatrices, and
eigenvalues for these equations as well. However, the techniques we have applied
here may also be useful for other nonlinear differential equations such as the
Thomas-Fermi equation $y''(x)=[y(x)]^{3/2}/\sqrt{x}$, which is posed as a
boundary-value problem satisfying the boundary conditions $y(0)=1$ and $y(\infty
)=0$. The solution to this problem is {\it unstable} with respect to small
changes in the initial data; if the initial slope $y'(0)$ is varied by a small
amount, the solution develops a spontaneous singularity at some positive value
$a$. A leading-order local analysis suggests that this singularity is a
fourth-order pole of the form $400(x-a)^{-4}$. However, this singularity is not
a pole. Indeed, a higher-order local analysis indicates that there is a
logarithmic-branch-point singularity at $x=a$ as well and thus the solutions to
the Thomas-Fermi equation live on multisheeted Riemann surfaces. It would be
interesting to see if our work on nonlinear eigenvalue problems extends beyond
meromorphic functions.

\end{document}